\pdfoutput=1
\documentclass[sn-mathphys]{sn-jnl}

\jyear{2023}%

\theoremstyle{thmstyleone}%
\theoremstyle{thmstyletwo}%

\newcommand{\appropto}{\mathrel{\vcenter{
  \offinterlineskip\halign{\hfil$##$\cr
    \propto\cr\noalign{\kern2pt}\sim\cr\noalign{\kern-2pt}}}}}

\theoremstyle{thmstylethree}%
\usepackage{amsmath}
\usepackage{lineno}
\usepackage{subfig}
\usepackage{xcolor}
\usepackage{rotating}

\begin{document}

\title[Physics Constrained Deep Learning for Ptychographic Reconstruction]{Physics Constrained Unsupervised Deep Learning for Rapid, High Resolution Scanning Coherent Diffraction Reconstruction}

\author*[1]{\fnm{Oliver} \sur{Hoidn}}\email{ohoidn@slac.stanford.edu}

\author[1]{\fnm{Aashwin Ananda} \sur{Mishra}}\email{aashwin@slac.stanford.edu}

\author[1]{\fnm{Apurva} \sur{Mehta}}\email{mehta@slac.stanford.edu}

\affil[1]{SLAC National Accelerator Laboratory, Menlo Park, California, USA}

\abstract{By circumventing the resolution limitations of optics, coherent diffractive imaging (CDI) and ptychography are making their way into scientific fields ranging from X-ray imaging to astronomy. Yet, the need for time consuming iterative phase recovery hampers real-time imaging. While supervised deep learning strategies have increased reconstruction speed, they sacrifice image quality. Furthermore, these methods' demand for extensive labeled training data is experimentally burdensome. Here, we propose an unsupervised physics-informed neural network reconstruction method, PtychoPINN, that retains the factor of 100-to-1000 speedup of deep learning-based reconstruction while improving reconstruction quality by combining the diffraction forward map with real-space constraints from overlapping measurements. In particular, PtychoPINN significantly advances generalizability, accuracy (with a typical 10 dB PSNR increase), and linear resolution (2- to 6-fold gain). This blend of performance and speed offers exciting prospects for high-resolution real-time imaging in high-throughput environments such as X-ray free electron lasers (XFELs) and diffraction-limited light sources.}

\maketitle

\section{Introduction }\label{sec1}
Coherent diffractive imaging (CDI) is a state-of-the-art imaging technique that uses diffraction from a coherent beam of light or electrons to reconstruct an image of a specimen without the need for optics. This microscopy technique is useful because it overcomes the lens aberrations of traditional lens-based imaging, thus allowing imaging at finer scales than previously possible. As a general approach CDI has found application in a broad range of settings, including nanoscale imaging through Bragg coherent diffractive imaging reconstruction (BCDI), X-ray ptychography, optical super-resolution, and astronomical wavefront sensing. \cite{dean2006phase, heintzmann2021answers, miao2015beyond}

The main challenge of CDI, known as the phase retrieval problem, originates from the fact that detectors only record the intensity (i.e., squared amplitude) of the diffracted wave, not its phase. The phase carries essential information about the illuminated real-space object, and its loss stops the direct calculation of an image from the recorded diffraction. A breakthrough development over two decades ago outlined an iterative approach to solve the inverse problem of phase retrieval and made CDI image reconstruction possible \cite{miao1999extending}. Unfortunately, these iterative algorithms are slow and typically computationally expensive \cite{epie}, which consequently precludes CDI in high-throughput or in situ settings, such as at x-ray free electron lasers (XFELs). In addition, such methods often suffer from noise sensitivity and limitations in robustness.

A considerable body of literature focuses on applying deep learning (DL) to inverse problems, with significant success in employing neural networks (NNs) to solve the CDI phase problem much more rapidly than conventional iterative methods \cite{ratner2021recovering,yao2022autophasenn,chang2023deep}. Early efforts used supervised learning to train these deep learning models and achieved several-orders of magnitude speed improvements, accompanied by two major drawbacks: degradation in reconstruction quality and the need for large volumes of high-quality labeled training data. More recently, strategies such as incorporating the diffraction forward map into deep learning models have been introduced to eliminate the requirement for labels. 

Despite the promise of these developments, existing physics-informed neural networks (PINNs) for CDI have not explored physical priors or constraints beyond the diffraction forward map and discrete Fourier transform (DFT) sampling requirements, nor have they modeled any stochasticity of the relevant physics. Both these approaches introduce Inductive Biases in the learning algorithm \cite{mitchell1980need, baxter2000model}, enabling better generalization to test samples outside the explicit training set while also reducing the need for large corpora of training data. As an illustration, incorporation of physics based knowledge via hard constraints reduces the region in solution space to be explored and thus reduces the need for large training datasets \cite{baker2019workshop}. Similarly, incorporating uncertain or incomplete domain-specific knowledge via the use of probabilistic loss functions leads to improved convergence and generalizability \cite{baker2019workshop}. In this light, it is conceivable that incorporating additional physics information in the model architecture and developing principled probabilistic loss functions may enhance both accuracy as well as generalizability, while also maintaining the speed of prior supervised learning approaches. 

In this work, we develop the PINN computational approach to CDI, specifically for ptychographic reconstruction. Ptychography is defined as a conventional CDI technique with overlap, originally introduced to calculate the phase of Bragg reflections from crystals. It represents an amalgam between Scanning Transmission X-ray Microscopy (STXM) and Coherent Diffraction Imaging (CDI). Ptychography utilizes convolutions in Fourier space between an object's diffraction pattern and the transform of the illumination function. This involves the measurement of manifold diffraction patterns via the scanning of an X-ray probe over the specimen, while using the overlap between adjacent illuminations for over-determination. In this light, it is often also referred to as scanning CDI. When coupled with reconstruction approaches, this leads to robust computational imaging. 

In this study, we integrate three novel elements for Machine Learning based approaches to ptychography: unsupervised training using the diffraction forward map, additional physics-based constraints informed by the ptychography setup, and an explicitly probabilistic treatment of photon counting (Poisson) statistics. We find that this unique combination of model features merges the advantages of a standard PINN approach, namely speed and unsupervised training, with significantly better generalizability and reconstruction accuracy than other NN-based solvers, including physics-informed approaches.

\section{Methods, Models \& Tests}
\subsection{Approach}

Physics-based CDI reconstruction methods are accurate because they invert the physically correct forward map of far field diffraction, making them capable of finding the optimal solution, in principle, for any input. However, due to the difficult nonconvex optimization problem, these methods require computationally expensive iterative solution schemes. Additionally, iterative reconstruction is oblivious to regularities in the input data, so each new diffraction signal requires computation commencing from scratch. In contrast, NN-based reconstruction methods take a different approach: they may not incorporate domain knowledge of the diffraction physics but instead rely on a large amount of training data to fit a flexible black box model from scratch. The inductive biases introduced in the learning algorithm are chiefly from the choice of the model architecture. For instance, the use of convolutional and pooling layers renders the final model approximately translation invariant. The lack of domain knowledge-based inductive biases and physical consistency-enforcing constraints in conventional NN-based methods cause them to have reduced accuracy and generalization with respect to the underlying physics, although they may capture particular data regularities well. Additionally, conventional neural networks are trained via supervised learning approaches. Thus, they require large corpora of labeled data for training, and the limited diversity of the training data brings in bias. Nonetheless, the single forward pass nature of NN-based methods makes them intrinsically rapid at inference time.

From this perspective, physics-informed neural networks (PINNs) attempt to unite the best of both worlds\footnote{Archetypal PINNs utilize a soft-constraint on the solution space by using the residual of the governing Partial Differential Equation as a regularization term. Our PINN model incorporates domain physics information in the architecture of the model.}. By strongly constraining a neural network model's hypothesis space to exclude parameter combinations that generate unphysical solutions, we can predispose a model towards physically correct solutions while also considerably reducing the need for training data. As a concrete starting point, defining the model's loss function over the forward-mapped (i.e., far field-diffracted) NN output -- instead of the immediate NN output -- forces the NN to learn diffraction physics rather than merely fit \emph{a priori} arbitrary input/output pairs. This is the foundation for prior PINN approaches for unsupervised CDI reconstruction \cite{yao2022autophasenn, ratner2021recovering}.

\subsubsection{Map formulation}
With this background in place, we develop a new approach for ptychographic reconstruction based on a combination of NN layers, explicit constraints, and the forward map of far-field diffraction.

To begin, the reconstruction problem requires approximating a mapping $G: X \rightarrow Y$ from the diffraction/reciprocal-space domain $X$ to the real-space domain $Y$. Because we wish to avoid supervised training, we rely on an autoencoder formalism that composes $G$ with a second mapping $F: Y \rightarrow X$. The output of the autoencoder is then $\hat{x} = F(G(x))$, where $x$ is the (complex) diffracted wave field. \footnote{The measured diffraction intensity is therefore $I \appropto \lvert x \rvert ^2 $.}

In prior PINN approaches to CDI reconstruction, $F$ is typically the forward map of far-field diffraction. Here, we instead start by subdividing $F$ into two parts: a \emph{constraint} map $ F_c: Y \rightarrow Y$ and a diffraction map $ F_d: Y \rightarrow X$, such that $F(Y) = F_d(F_c(Y))$. In most practical settings, diffraction amplitudes are centrosymmetric even for non-centrosymmetric objects. To break this centrosymmetry and make the inversion well-posed, $F_c$ imposes real-space constraints derived from overlapping diffraction measurements. $F$ depends on parameters $\theta$ of the experimental geometry (including the probe position), so we relate it to a functional $\mathbf{F}$, such that $F = \mathbf{F}(\theta)$. Notably, this formulation omits explicit dependence on the probe illumination function $P(r)$ of the diffraction map $F_d$ because we assume $P(r)$ to be known. In the analysis of simulated data we must therefore provide the ground truth $P(r)$ to the model, while for experimental data it is necessary to estimate $P(r)$ using an iterative solver such as ePIE.

In our concrete implementation the elements of a set of training samples $\{x_i\}$ have shape $64 \times 64 \times 4$, corresponding to four diffraction patterns measured at neighboring scan point coordinates. We denote a particular diffraction image with index $k$ $(k \in \mathcal{K} = \{0, 1, 2, 3\}$) as $x_i^k$. Each $x_i$ is paired with its matching 2D Euclidean probe coordinates, $r_i$ ($r_i \in \theta$). Correspondingly, the reconstruction $y_i = G(x_i)$ is a 3D tensor of shape $32 \times 32 \times 4$. (The factor-of-two difference in size between $x_i$ and $y_i$ satifies the oversampling requirement for invertibility of the discrete Fourier transform \cite{miao2000oversampling}). Finally, to distinguish between the two real-space representations we let $\bar{y}_i = F_c(y_i)$. We denote the amplitude and phase of $\bar{y}$ as $A$ and $\phi$, respectively, throughout this paper.

The inverse map $G$ consists of an encoder-decoder architecture with a similar structure to PtychoNN and $F_d$ is mostly defined by diffraction physics. The element of most interest is $F_c$, which we detail below.

\subsubsection{Real-space constraints}\label{sec_constraints}
In CDI reconstruction the phase problem manifests as invariances of the diffraction amplitude to coordinate inversion and translation of the real-space object. In the case of scanning CDI this must be solved using real-space constraints based on overlapping measurements.

In order to detail how the real-space constraint map, $\mathbf{F_c}(r_i)$, is applied to $y_i$, we can break it down into two steps. Initially, we sum the individual $y_i^k$ to create a unified reconstruction $\hat{y}_i$ with a shape of $64 \times 64$:

\begin{equation} 
\hat{y}_i = \Sigma_{k \in \mathcal{K}} T(r_i^k - \mu_i, Pad2d(y_i^k))
\oslash
\Sigma_{k \in \mathcal{K}} T(r_i^k - \mu_i, Pad2d(\bf{1})), 
\label{eq:1}
\end{equation}

where $Pad2d$ denotes zero-padding a $32 \times 32$ tensor to $64 \times 64$ and $T(\delta r, y)$ is the translation of a 2D tensor $y$ by the vector $\delta r$. $\mu_i$ is the origin of a local coordinate system for the group of scan points and is normally set to their centroid position, for convenience. $\oslash$ represents elementwise division, while $\bf{1}$ stands for a $32 \times 32$ tensor composed entirely of ones. Therefore, the divisor $\Sigma_{k \in \mathcal{K}} T(r_i^k - \mu_i, Pad2d(\bf{1}))$ has the role of a normalizing tensor.

The purpose of this first transformation is to construct a larger real-space image $\hat{y}_i$ that encompasses the entire solution region. The second, and arguably more critical, step enforces translational symmetry of the reconstruction via shifts in the relative probe position. This step also computes the exit wave, illuminating the shifted 2D object $\hat{y}_i$ with $P(r)$ and transforming it into a 3D object that matches the required input format of $F_d$:

\begin{equation}
\bar{y}_i^k = T(\mu_i - r_i^k, \hat{y}_i) P(r) \approx O(r) P(r - r_i^k),
\label{eq:2}
\end{equation}

Here, $O(r)$ represents the ground-truth object, with the domain of its discrete approximation (right hand side) confined to a $32 \times 32$ patch centered at $r_i^k$.

We can compare this approach to iterative scanning CDI reconstruction schemes, which enforce real-space constraints in a different way. In methods such as ePIE, the optimization loop is organized around alternating error corrections in real and reciprocal space. In the `backpropagation' step of ePIE, the error in the reconstructed diffraction amplitude for one measurement -- equivalent to our formulation's $\lvert x_i^k \rvert - \lvert \hat{x}_i^k \rvert $ -- induces an update of the $O(r)$ guess via inverse DFT. In the training of PtychoPINN, errors from the $\vert \mathcal{K} \vert$ individual patterns are converted into model parameter updates instead of reconstruction updates, and this is done in parallel instead of sequentially.

\subsubsection{Probabilistic output and loss function}

%

We choose the model output to be a collection of independent Poisson distributions parameterized by the forward-mapping of the final-layer output of the NN. This reproduces the correct photon-counting statistics of the diffraction measurement and allows us to calculate a likelihood, with respect to the Poisson parameters, for the distribution of per-pixel detected photon counts. The negative log over these Poisson likelihoods is \emph{a priori} a more principled loss function than the typical choice of mean absolute average (MAE) deviation between target and predicted pixel values.

Explicitly, the loss function is

$$
Loss(x, \lambda(\hat{x})) = \sum_{i,j,k}\log f_{\text{Poiss}}(x_{ijk}^2;\lambda_{ijk})
$$

where $\lambda_{ijk}(\hat{x}) = \hat{x}_{ijk}^2$, $x_{ijk}^2$ is the number of photons detected in a single pixel (its square root $x_{ijk}$ is the associated target amplitude), $\lambda_{ijk}$ is the matching final-layer output of the CNN, $i$ and $j$ index the detector coordinates, and $k$ indexes separate images within a diffraction set. 

For the above probabilistic formulation of the data, model output, and loss function to be self-consistent it is necessary for the units of the diffraction pixel values to be (unscaled) photon counts. A typical diffraction intensity is $10^9$ photons per exposure whereas the magnitude of activations within the NN -- including, in particular, the reconstructed real-space amplitude -- should be of order unity. To invertibly scale the input (and output) we define a global normalization parameter that we initialize using a simple heuristic based on the mean photon count of images in the training dataset and unitarity property of the Fourier transform.   This normalization parameter can optionally be either fixed or optimized during training. 

\subsection{PtychoPINN architecture}
\begin{figure}[h]%
\centering
\includegraphics[width=0.9\textwidth]{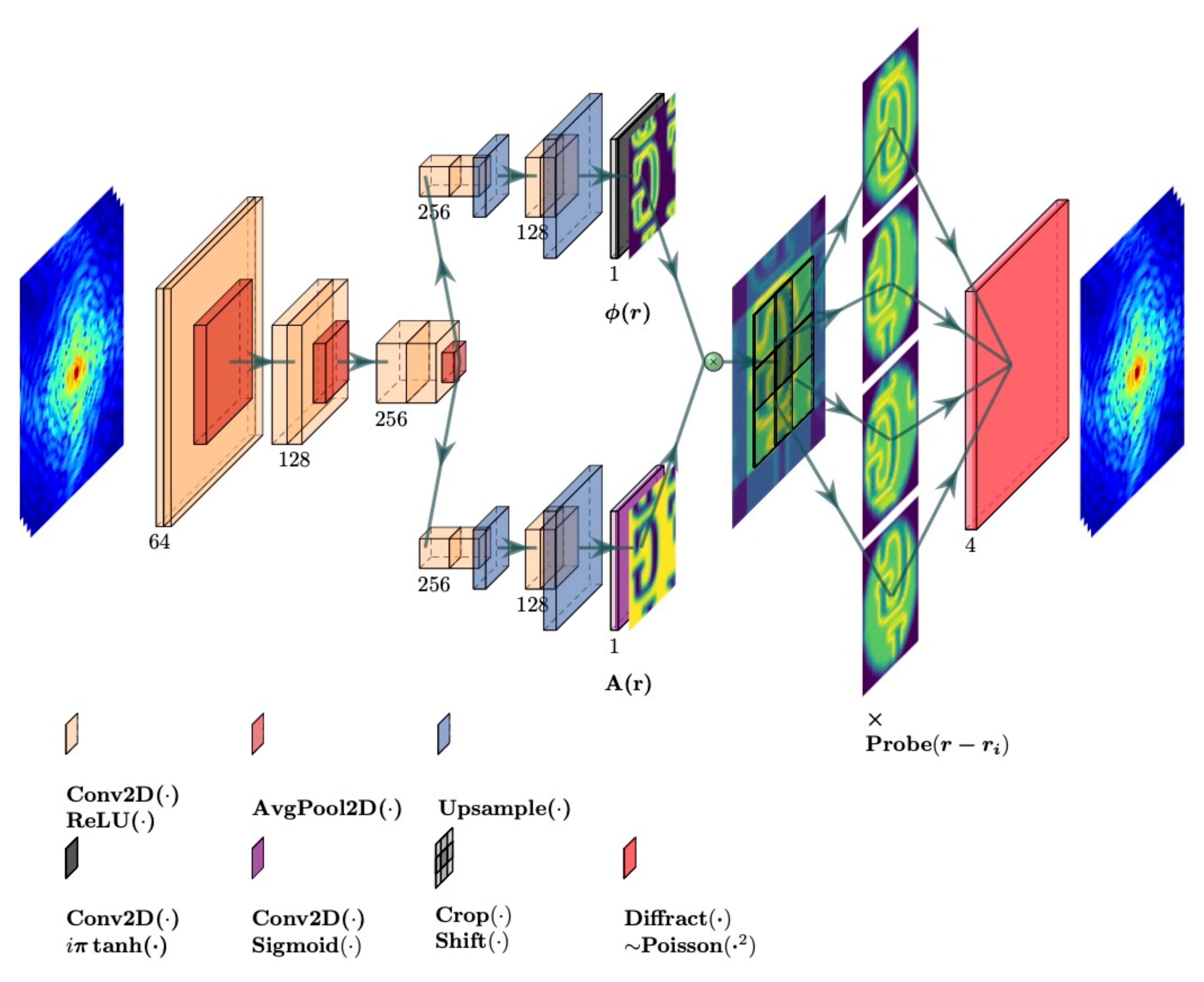}
\caption{Neural network architecture and training configuration of the PtychoPINN model.}\label{diagram}
\end{figure}

With the approach now laid out, we introduce its implementation in the form of the deep learning framework PtychoPINN. As illustrated in Fig. \ref{diagram}, PtychoPINN uses an autoencoder architecture that incorporates 2D convolutional, average pooling, and upsampling layers in addition to custom layers that scale the network input (and output) and apply the constraint transformations (\ref{eq:1}) and (\ref{eq:2}). All convolutional layers use rectified linear unit (ReLU) activations, and as in the BCDI model AutophaseNN we incorporate sigmoid and tanh activations to limit the domain of the phase and amplitude of $\bar{y}_i$ to $[0, 2 \pi]$ and $[0, 1]$, respectively. \cite{yao2022autophasenn}

\subsection{Data generation and training}\label{data}

\begin{figure}
    \centering
    {{\includegraphics[width=12cm]{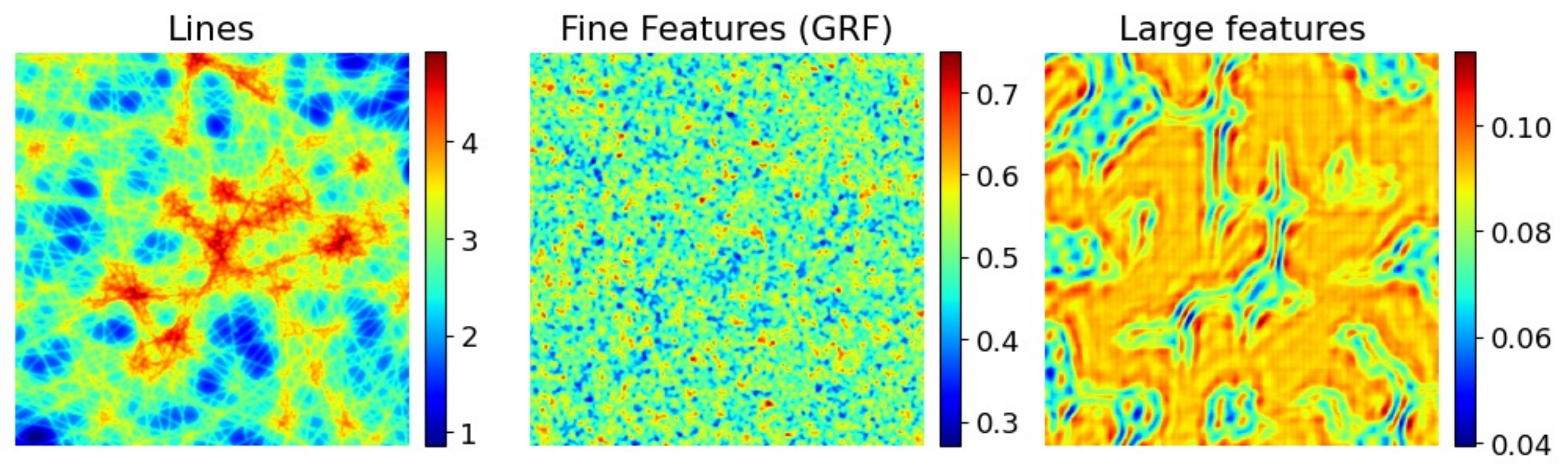} }}%
    \caption{\emph{Datasets.} Examples of amplitude images from three distinct dataset types used in this study. The `Lines' and `Fines Features' (Gaussian Random Field - GRF) datasets provide contrasting conditions of local symmetry (i.e., isotropy). The Gaussian random field process produces characteristic speckle, while `Lines' consists of sharp, oriented edges. `Large Features' is anisotropic but coarser than both `Lines' and `Fine Features'.}
    \label{fig:datasets}
\end{figure}

To prepare the model's training and evaluation data we begin with a collection of complex-valued images. We consider three types of images: simulated compositions of randomly-oriented lines (high aspect ratio features); samples with fine, isotropic features from a simulated Gaussian random field; and large features sampled from experimentally-derived phase and amplitude from x-ray ptychographic measurements of an etched tungsten test sample, which we retrieved from a publicly available dataset\cite{cherukara2020ai}. Representative samples of each dataset type are shown in Figure \ref{fig:datasets}, labeled respectively as Lines for randomly-oriented lines, GRF for Gaussian Random Field and `Large features' for the experimentally-derived data of \cite{cherukara2020ai}.

 Each of these datasets is characterized based on isotropy, sharpness, and the characteristic lengths in their real-space structure. Isotropy refers to uniformity of the statistical structure across all directions, i.e., invariance to rotation, reflection, and translation. The characteristic length signifies the spatial scale at which the correlation between two points diminishes. Sharpness refers to the granularity of the distinguishing features.
 
 As summarized in Table \ref{tab:comparison}, the GRF dataset is both isotropic and sharp, and has an extremely small characteristic length. In comparison, the `Lines' objects, which consist of overlays of oriented edge features, are sharp but lack isotropy. As is the case in anisotropic media, their characteristic lengths have a direction-dependence. Finally, the `Large features' dataset presents a distinct combination of anisotropy and coarseness. Like `Lines', it lacks isotropy, and the characteristic length of its features is large in two dimensions -- whereas that of `Lines' is large in only one dimension.

\begin{table}[h]
\centering
\renewcommand{\arraystretch}{1.5} 
\begin{tabular}{|l|c|c|c|}
\hline
\textbf{ } & \textbf{Isotropy} & \textbf{Sharpness} & \textbf{Characteristic Length} \\ \hline
\textbf{GRF} & Yes & Yes & Small \\ \hline
\textbf{Lines} & No & Yes & Mixed \\ \hline
\textbf{Large Features} & No & No & Large \\ \hline
\end{tabular}
\caption{Comparing the characteristics of the three dataset types used in this study.}
\label{tab:comparison}
\end{table}

For each real-space object dataset we simulate a collection of diffraction patterns corresponding to a rectangular grid of scan points on the sample and a known (complex-valued) probe function. Given the real-space object $O(r)$ and far-field diffraction forward map $F_d$, the simulated diffraction pixel values are random samples from $f_{\text{Poiss}}(F_d(O(r))^2)$, where $f_{\text{Poiss}}$ is the Poisson distribution.

For all object types except `Large features', each training dataset contains 49,284 diffraction patterns densely sampled from 9 simulated objects (5476 patterns per object), and the size of each simulated object is $392\times 392$ pixels. (The `Large features' dataset is limited to a single object with 16,100 diffraction patterns, corresponding to its smaller underlying experimental dataset.) The probe positions were sampled on a nested rectangular grid with a spacing of 8 pixels between solution regions and a spacing of 4 pixels probe positions within a solution region. (As defined in \ref{sec_constraints}, a solution region comprises a $2\times 2$ local grid of overlapping diffraction measurements.) To generate diffraction from an object patch we assume the ground truth probe illumination function $P(r)$ and take as a simulation parameter the expected number of photons incident on-detector, which we set to $10^9$ photons for all simulated data presented in this paper. 

To train the model parameters we use the Adaptive Moment Estimation (ADAM) optimizer with an initial learning rate of 0.001.\cite{kingma2014adam} We train for 50 epochs and with a batch size of 16, which takes approximately 10 minutes on an Nvidia RTX 3090 GPU.

\section{Results and Discussion}

\subsection{Numerical experiments}

\begin{sidewaysfigure}%
    \centering
    \subfloat[\centering Ground truth $A$]{{\includegraphics[width=6cm]{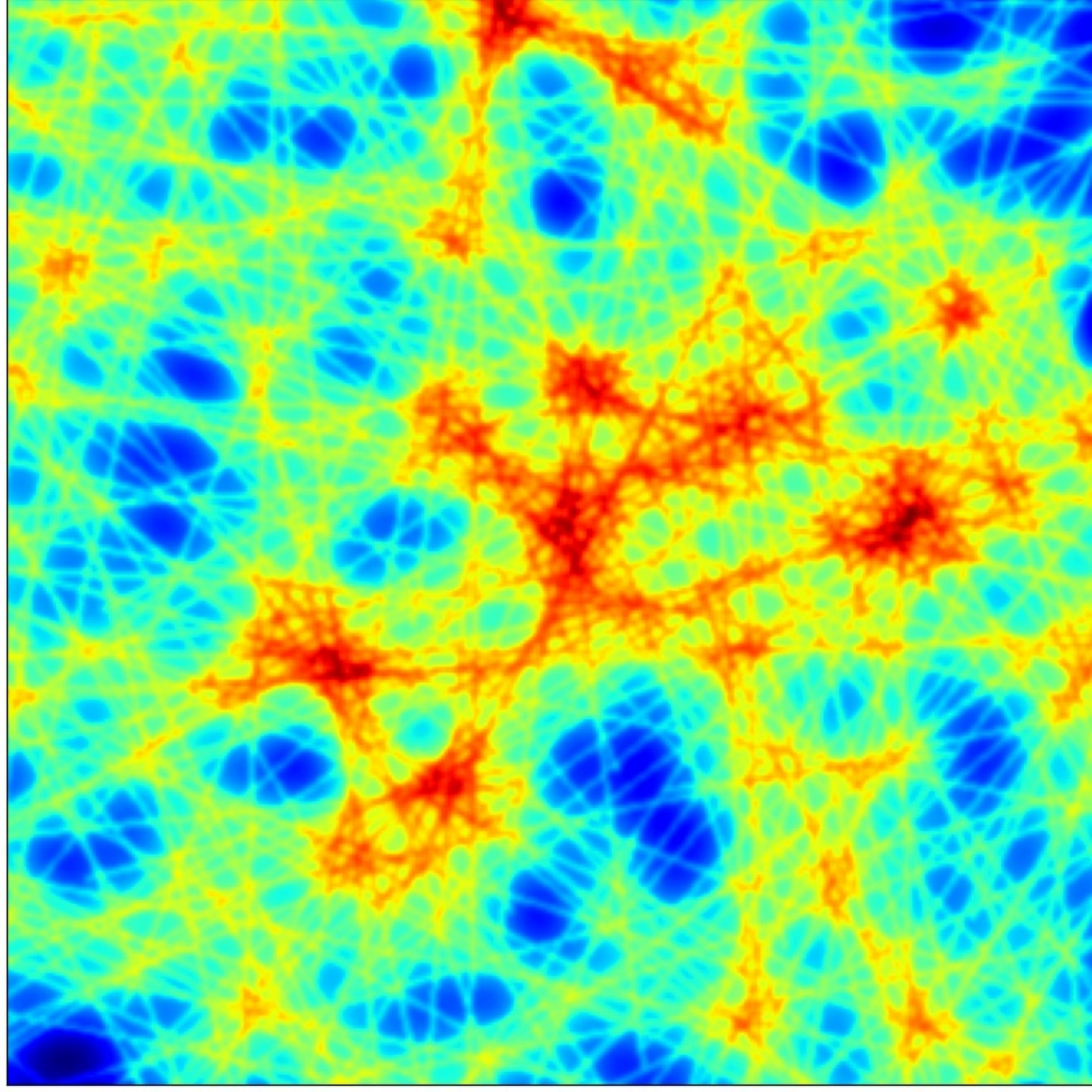} }}%
    \hspace*{0cm}
    \subfloat[\centering PtychoNN $A$]{{\includegraphics[width=6cm]{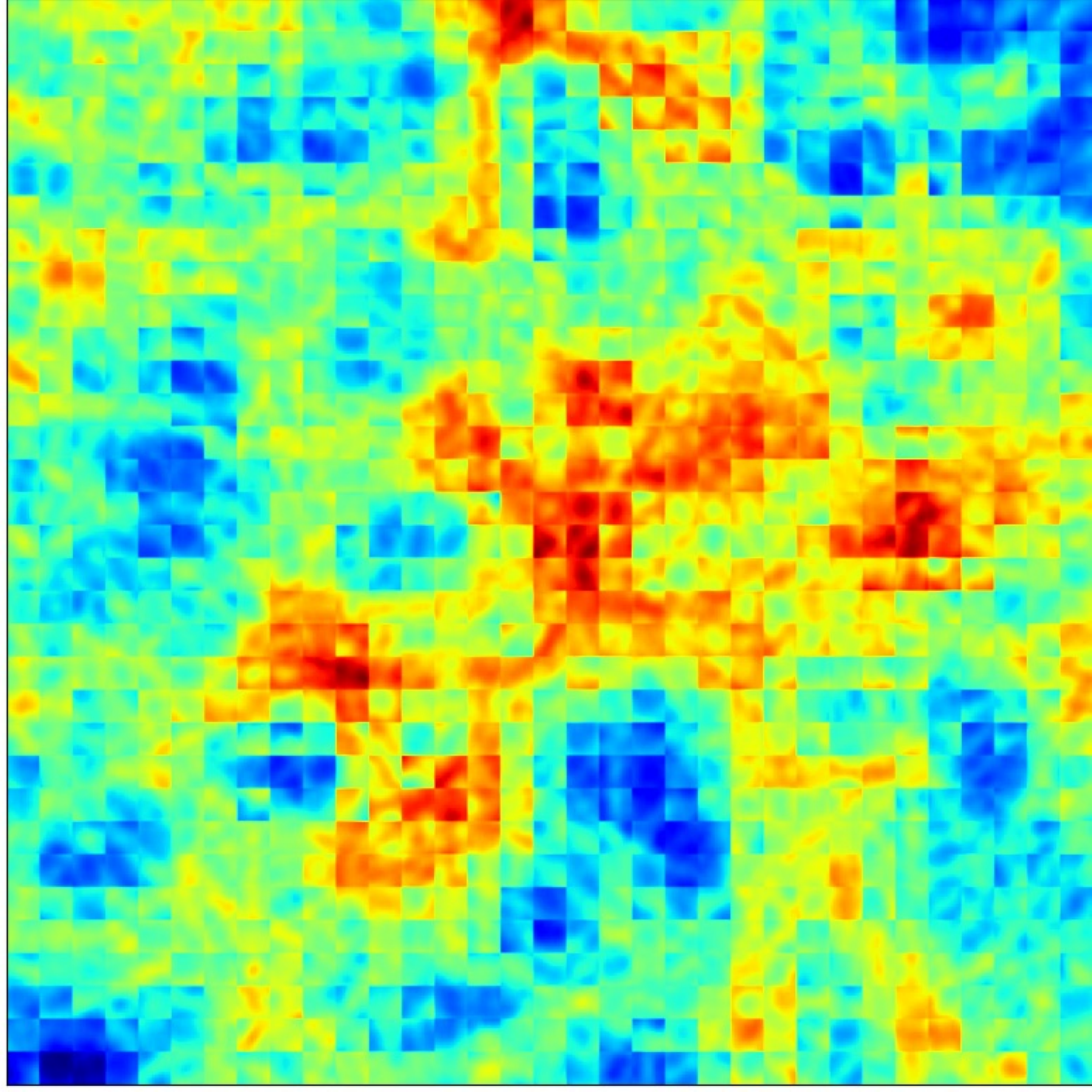} }}%
    \hspace*{0cm}
    \subfloat[\centering PtychoPINN $A$]{{\includegraphics[width=6cm]{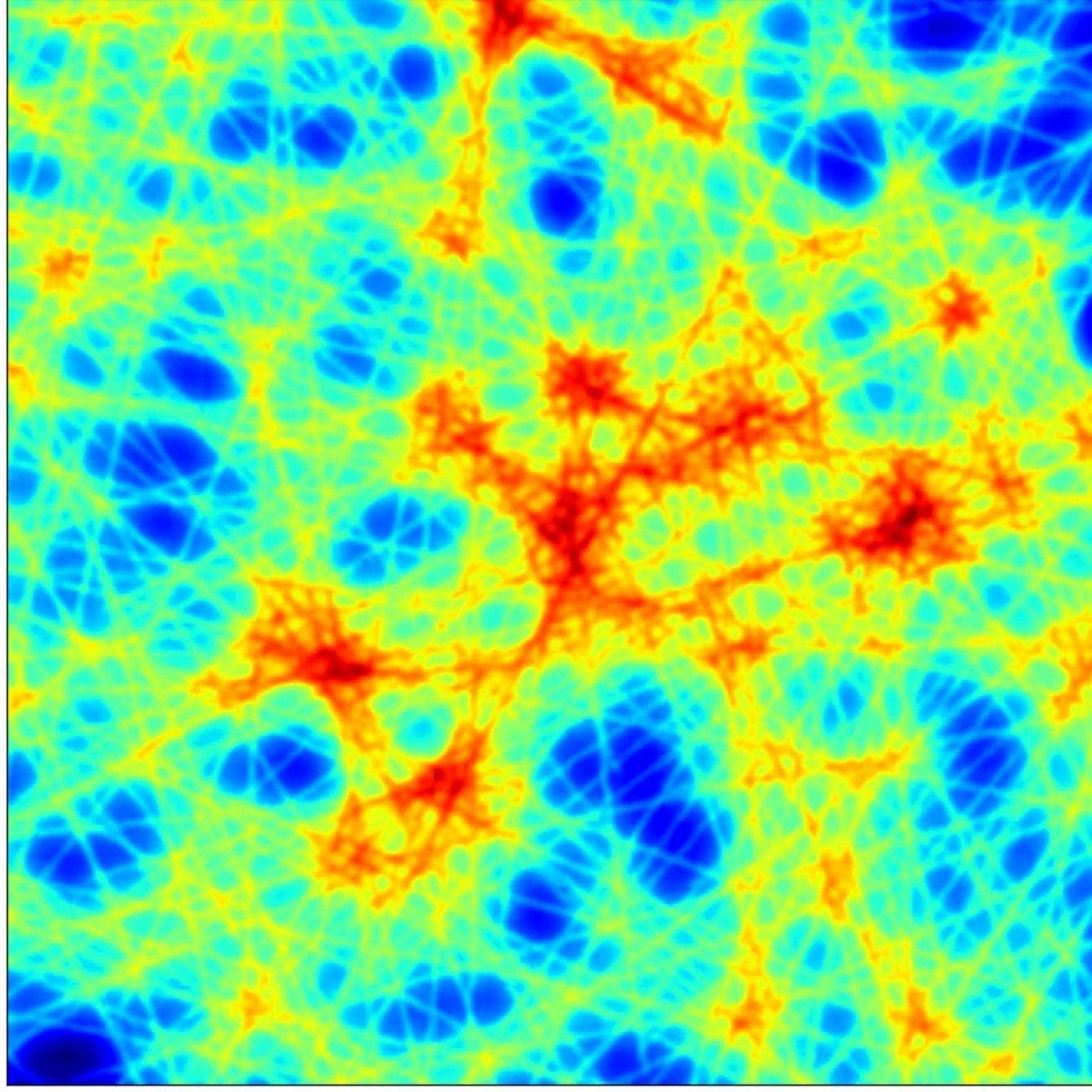} }}%
    \caption{Representative Amplitude reconstruction comparison for the Lines dataset.}%
    \label{fig:sim_comparison}%
\end{sidewaysfigure}

\begin{table}[h]
\begin{center}
\caption{Three reconstruction metrics for the baseline supervised-learning model (PtychoNN) and PtychoPINN, repeated for datasets with contrasting asymmetry and feature sharpness: Lines, GRF and `Large features'. For each combination of dataset (top column header) and reconstruction metric, the best amplitude and phase reconstructions are bolded. Lines, with its sharp, asymmetric features, yields the largest improvements in reconstruction of the amplitude image (bottom left of table). The baseline model recovers large amplitude features quite well (rightmost column), but PtychoPINN is more robust in its phase reconstruction across all datasets.  
}\label{tab1}
\begin{tabular}{p{2cm}l|ll|ll|ll}
\toprule
& \multicolumn{1}{c}{} & \multicolumn{2}{c}{Lines \footnotemark[1]} & \multicolumn{2}{c}{GRF} & \multicolumn{2}{c}{Large features}\\
\midrule
\bf{Model} & \bf{Metric} 
& $\phi$ & $A$
& $\phi$ & $A$
& $\phi$ & $A$ \\
\midrule
PtychoNN
& MAE & - & 0.201 & 0.0335 & 0.0153 & 0.219 & 0.0038 \\
& PSNR (dB) & - & 59.6 & 75.6 & 82.4 & 56.7 & 92.9 \\
& FRC50 ($\mathrm{pixel}^{-1}$) & - & 22.0 & 64.0 & 65.2 & 23.4 & 34.0 \\
\midrule
PtychoPINN
& MAE & - & \textbf{0.0473} & \textbf{0.0109} & \textbf{0.00507} & \textbf{0.149} & \textbf{0.00303} \\
& PSNR (dB) & - & \textbf{72.6} & \textbf{85.2} & \textbf{91.9} & \textbf{60.6} & \textbf{95.0} \\
& FRC50 ($\mathrm{pixel}^{-1}$) & - & \textbf{165.4} & \textbf{171.5} & \textbf{171.3} & \textbf{93.7} & \textbf{38.7} \\
\midrule
\end{tabular}
\end{center}
\end{table}
\footnotetext[1]{We omit metrics for phase images of the `Lines' datasets, as these are all zero-valued.}

Figure \ref{fig:sim_comparison} compares the reconstruction of artificially generated objects—specifically, those from the `lines' data type in Figure \ref{fig:datasets}—using PtychoPINN and the supervised learning baseline, PtychoNN. PtychoPINN exhibits minimal degradation in the real-space amplitude and phase. In contrast, the supervised learning baseline experiences significant blurring. The amplitude Fourier ring correlation at the 50 percent threshold (FRC50)--a useful measure of linear resolution--is $165.4~\mathrm{pixel}^{-1}$ for PtychoPINN, a notable improvement to the baseline's $22.0~\mathrm{pixel}^{-1}$. Additionally, PtychoPINN's peak signal-to-noise ratio (PSNR), a standard super-resolution metric, is 13 dB higher.

We further assessed the model's performance across diverse data by including Gaussian random field (GRF) generated images and the `Large features' object derived from experimental measurements (Table \ref{tab1}). For each image type, PtychoPINN significantly outperforms the baseline PtychoNN in both phase and amplitude, as per the MAE metric. However, enhancements in PSNR and FRC50 were comparatively modest for images featuring larger (`Large Features') or more equiaxed (`GRF') features, when juxtaposed with the improvements observed for the lines dataset.

\begin{sidewaysfigure}%
    \centering
    \subfloat[\centering Ground truth $\phi$]{{\includegraphics[width=5.5cm]{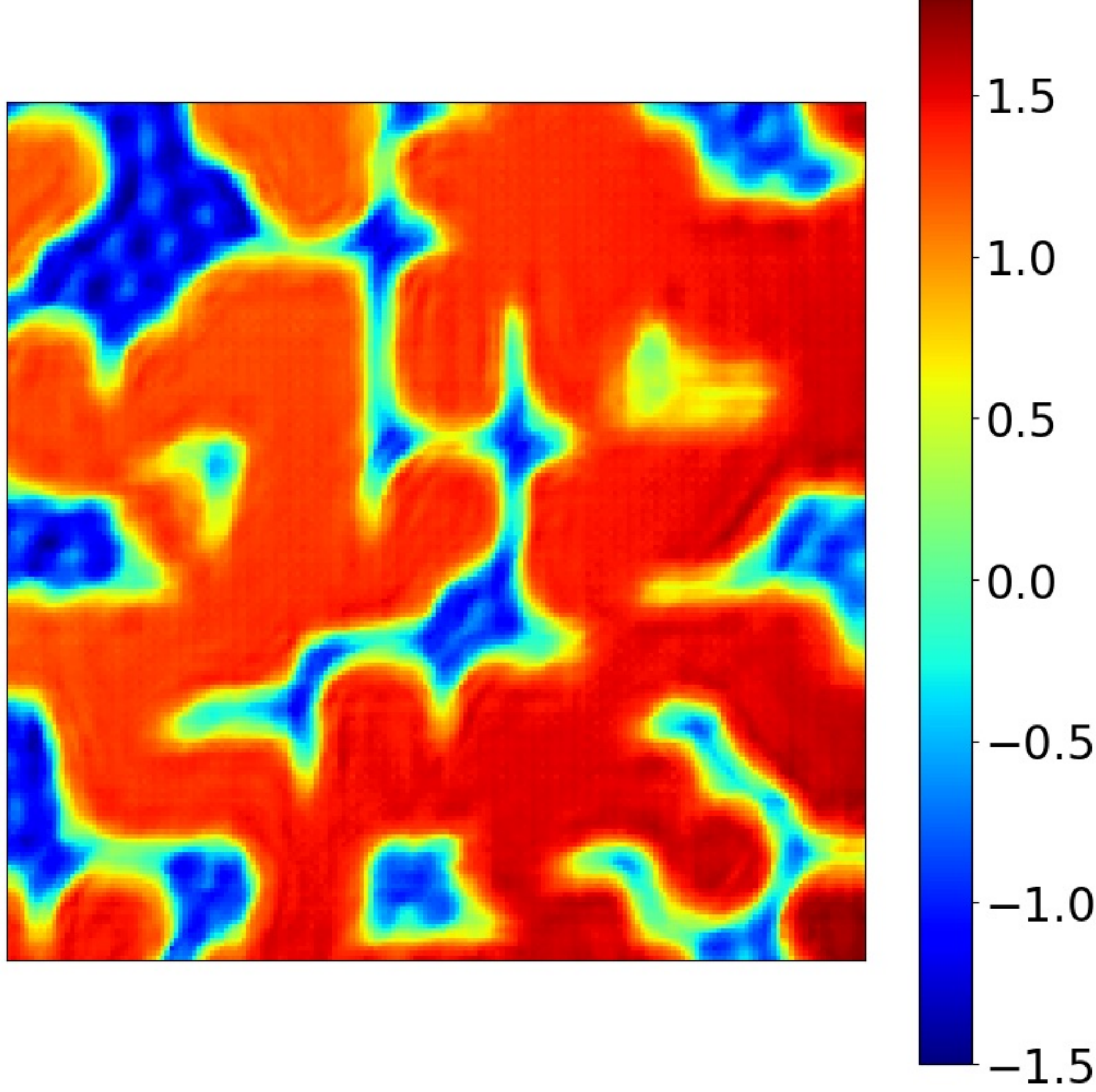} }}%
    \subfloat[\centering PtychoNN $\phi$]{{\includegraphics[width=5.5cm]{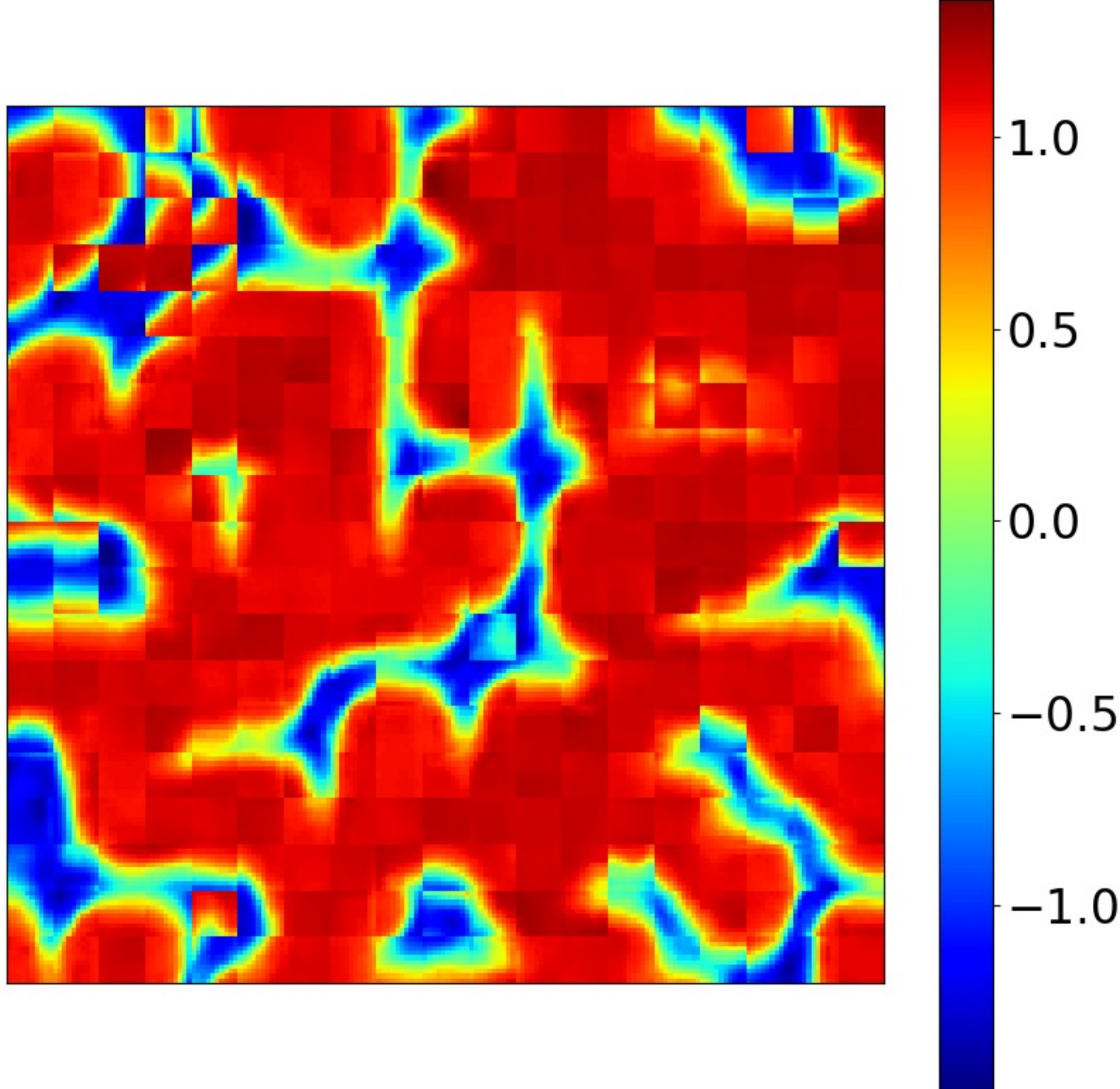} }}
    \subfloat[\centering PtychoPINN $\phi$]{{\includegraphics[width=5.5cm]{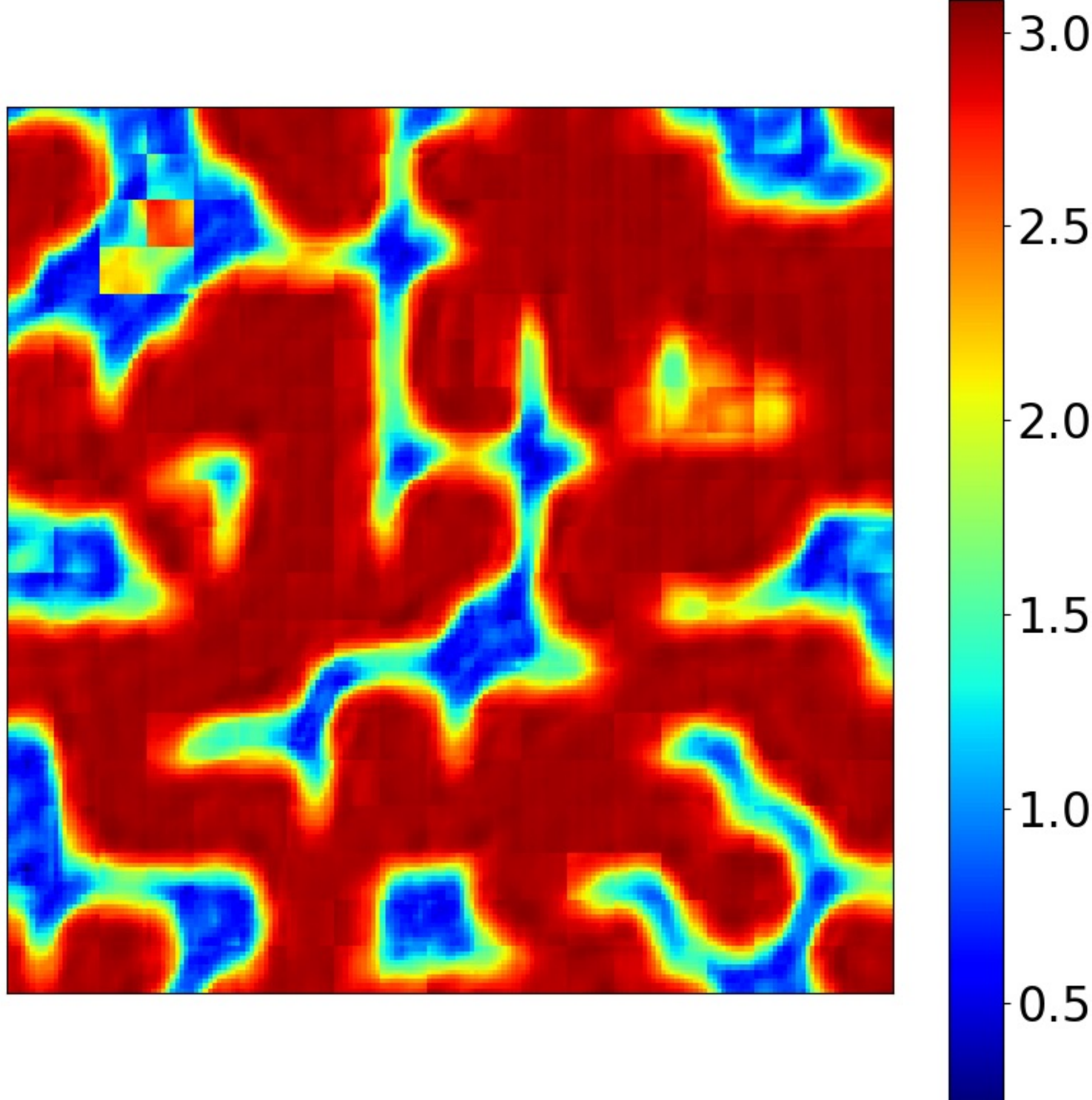} }}
\vfill
    \subfloat[\centering Ground truth $A$]{{\includegraphics[width=5.5cm]{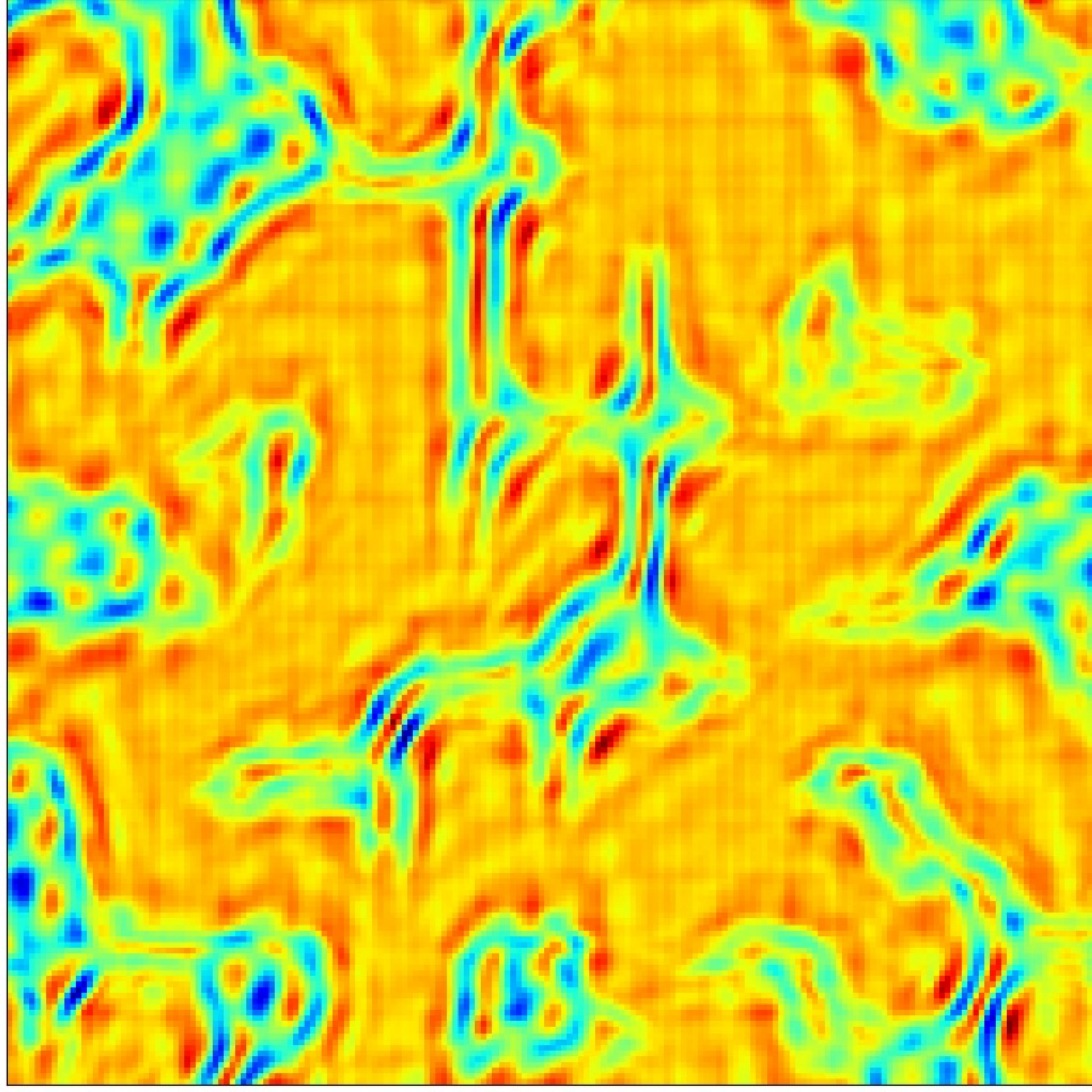} }}%
    \subfloat[\centering PtychoNN $A$]{{\includegraphics[width=5.5cm]{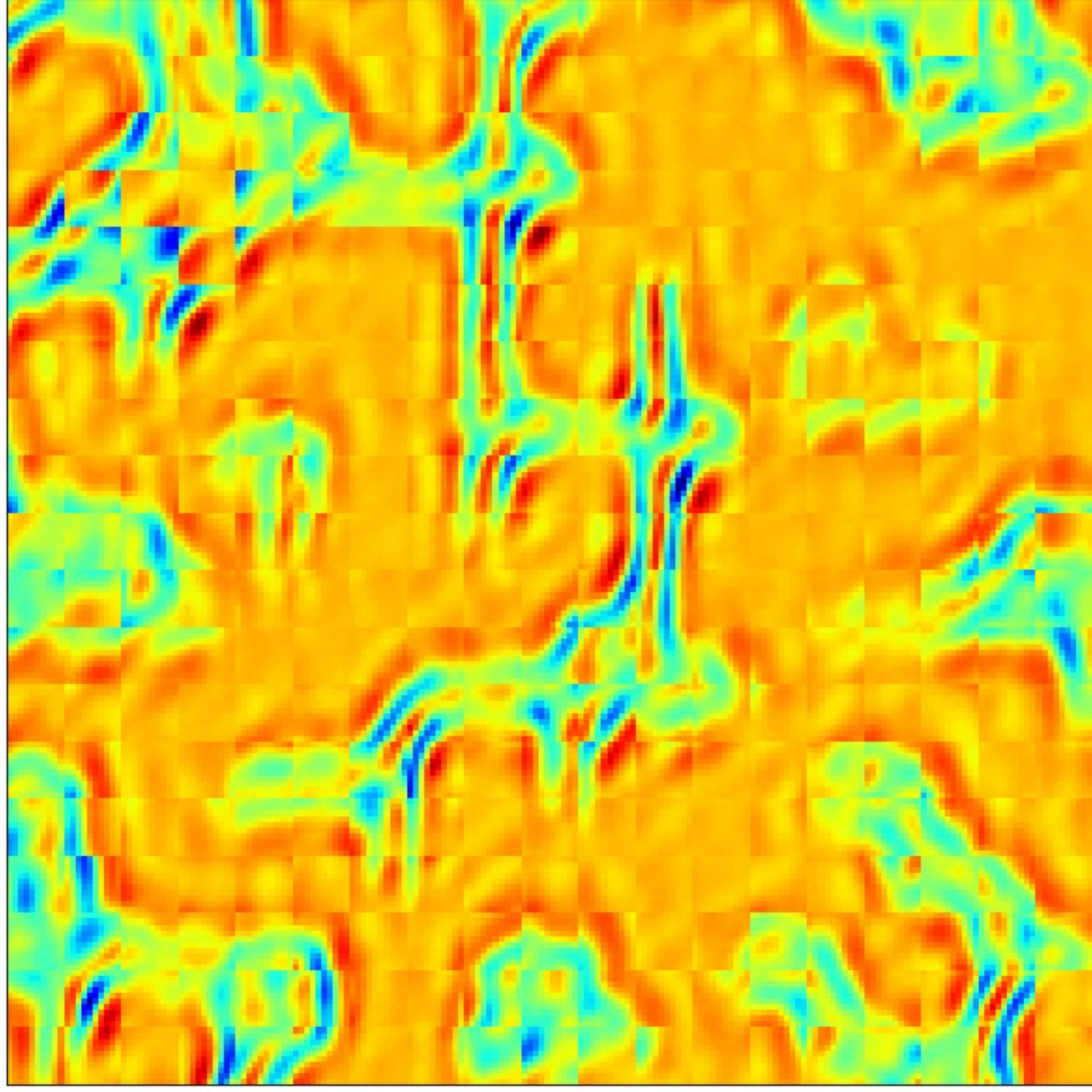} }}
    \subfloat[\centering PtychoPINN $A$]{{\includegraphics[width=5.5cm]{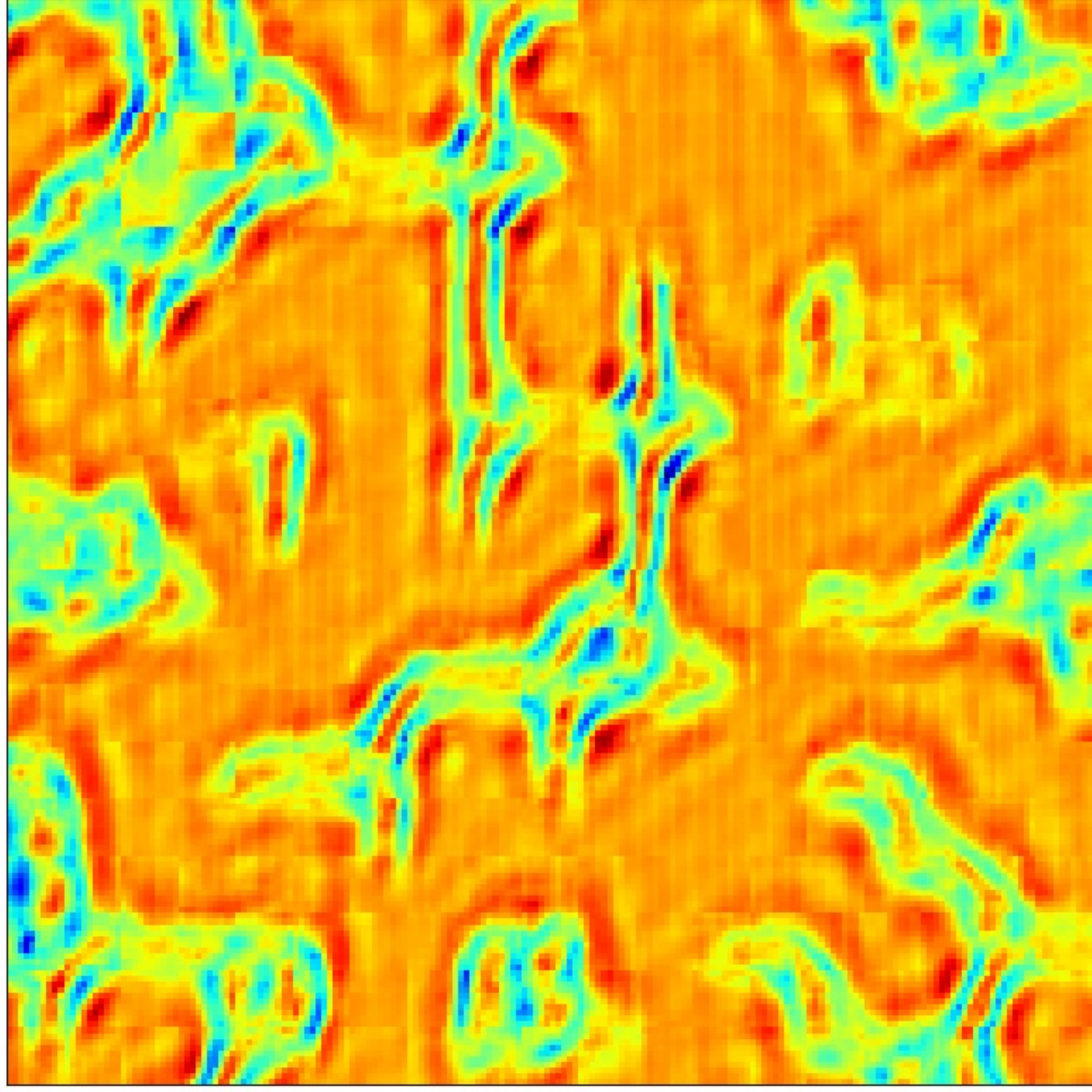} }}%
    \caption{ \emph{Amplitude and phase reconstruction.} Comparison of PtychoPINN to a supervised learning method, PtychoNN, on the `Large features' object. As the amplitude object lacks sharp features, PtychoNN reconstructs it nearly as well as PtychoPINN. However, it tends to invert the phase structure in low-amplitude regions.}
    \label{fig:exp_comparison_detailed}%
\end{sidewaysfigure}

As shown in Fig. \ref{fig:exp_comparison_detailed}, both PtychoPINN and PtychoNN provide good reconstructions of the `Large features' amplitude. However, the baseline model's phase reconstruction is less favorable, with an FRC50 of $23.4~\mathrm{pixel}^{-1}$ compared to PtychoPINN's $93.7~\mathrm{pixel}^{-1}$. In regions of the object with low scattering amplitude, phase reconstruction is less reliable than that of the amplitude under both models. However, PtychoPINN is more robust, producing significantly fewer artifacts in those regions compared to PtychoNN (Fig. \ref{fig:exp_comparison_detailed}(b) and (c)). Crucially, improving the estimation of the object in phase space not only results in a higher-quality final image, but also ensures the problem is well-posed \cite{millane1990phase} and leads to a unique solution.

The real-space image is the full, complex-valued $O(r)$ -- not only its phase or amplitude, as portrayed for convenience in Fig \ref{fig:sim_comparison} and elsewhere. A competent image reconstruction requires both the amplitude and phase of $O(r)$ and, moreover, the phase image sometimes proves more informative than the amplitude image, such as in the case of biological materials that have small variations in the differential scattering cross-section but good contrast in $\phi(r)$.


\subsubsection{Ablation study}

To better understand the specific aspects of the PtychoPINN architecture that contribute to its improved performance, we conducted an ablation study evaluating the impact of PINN-based training and real-space constraints. The comparison of reconstruction accuracy (MAE), resolution (Fourier ring correlation, FRC), and peak signal-to-noise ratio (PSNR) in  Supplementary Table 1  suggests that the combination of overlap constraints and PINN structure is the most significant factor in improving performance. \emph{Combination} is an important qualifier, as neither the PINN architecture nor the overlap constraints yields a substantial improvement in reconstruction accuracy when used individually.

\subsubsection{Out-of-distribution generalization}
To complement these standard evaluation metrics we tested PtychoPINN's out-of-distribution generalization. For this purpose, we extracted two distinct $392 \times 392$ patches from a larger image and used $26,896$ diffraction patterns, simulated from just one patch for training both the PtychoPINN model and the baseline model (Figure \ref{fig:gen_detailed}).

In order to provide a rigorous test of the models' out-of-distribution performance, we intentionally selected two image patches with considerable divergence in their features. This choice of markedly different images creates a more stringent test of out-of-distribution behavior than the other datasets that we have presented.

Evaluating the models' ability to reconstruct both in-distribution and out-of-distribution patches provides insight into their generalization capacity. As shown in Fig. \ref{fig:gen_detailed}, a comparison between panels (c) and (f) (representing PtychoPINN) and panels (b) and (e) (representing the baseline - PtychoNN - model) indicates that PtychoPINN experiences a less severe drop in out-of-distribution reconstruction quality, implying better generalizability compared to the baseline. This observation is further backed by numerical comparison (Table \ref{tab2}). As expected, both models perform less well out-of-distribution, with a more pronounced decline in the FRC50 (resolution) metric than the PSNR metric. However, PtychoPINN outperforms the baseline model, with almost double the resolution.

\subsubsection{Performance comparison to Iterative Solvers}
The reconstruction of a dataset of $1024$ diffraction images using the trained model takes $0.3$ seconds. In comparison, a CPU implementation of ePIE (without position correction) completes in $165$ seconds. 

\subsection{Discussion}
In summary, we present an unsupervised learning approach for scanning coherent diffraction imaging (CDI) that integrates real-space constraints with a physics-informed neural network (PINN) architecture (PtychoPINN) to yields substantial improvements in reconstruction accuracy and resolution while preserving the inherent speed of the previous NN-based approaches. 

In the ensuing discussion, our primary objective is to highlight features of PtychoPINN  that would be most important to a domain scientist intending to apply it. When possible, we will also take the perspective of a machine-learning researcher to understand \emph{why} PtychoPINN exhibits these characteristics. We specifically concentrate on three essential attributes: unsupervised training, generalizability, and the interplay between resolution and accuracy.

\subsubsection{Unsupervised training} 
The use of unsupervised training in our approach offers three important advantages over supervised learning methods in an experimental setting. Firstly, supervised methods require labeled data, placing the onus on the experimenter to first gather or simulate numerous ptychographic datasets, and then painstakingly reconstruct the real-space images using iterative methods. Only then can a faster neural network-based method be trained.

Secondly, the demand for labeled data in supervised learning restricts the ability to retrain or update the model 'on-the-fly.' Unsupervised training, on the other hand, allows the use of training samples drawn from the same distribution as the test-time samples. This reduces the risk of any issues related to out-of-distribution generalization that the model may otherwise encounter, which aids robustness -- particularly in real-time settings.

Lastly, supervised learning methods can introduce biases due to the lack of diversity in the image types within the labeled training data, which in turn reduces the model's generalizability. We delve deeper into addressing this issue in the following sections.

\subsubsection{Generalizability} \label{sec_generalization}

\begin{sidewaysfigure}%
    \centering
    \subfloat[\centering Ground truth]{{\includegraphics[height=4cm]{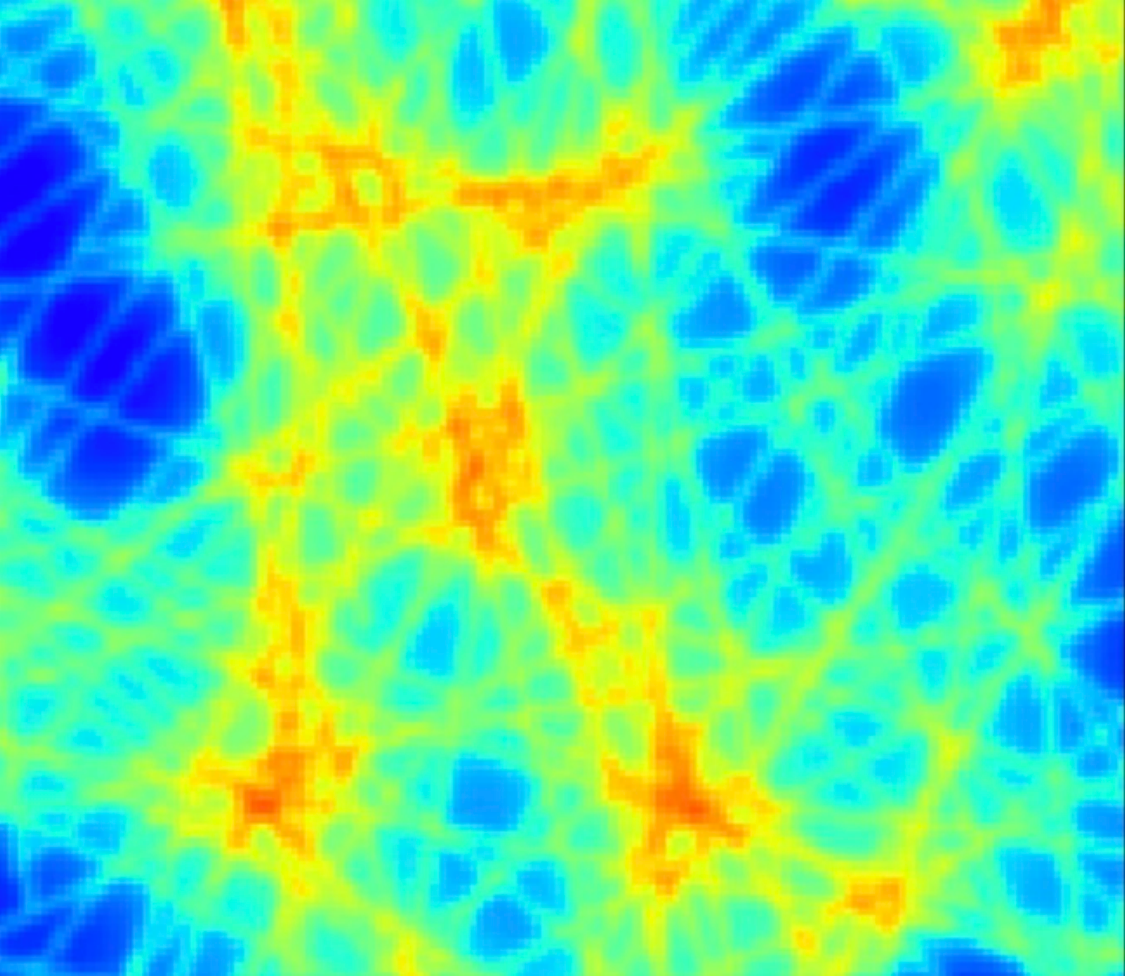} }}%
    \subfloat[\centering PtychoNN]{{\includegraphics[height=4cm]{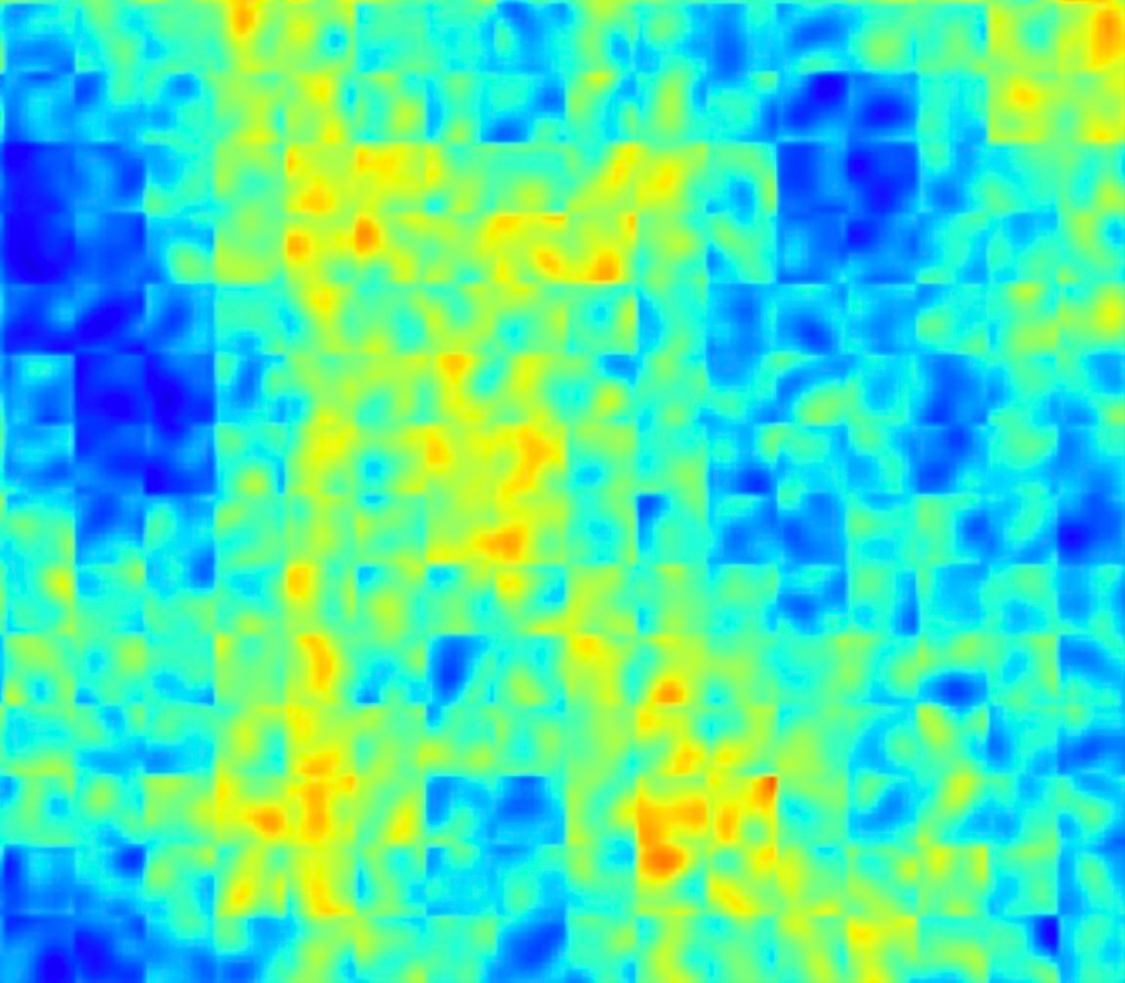} }}%
    \subfloat[\centering PINN]{{\includegraphics[height=4cm]{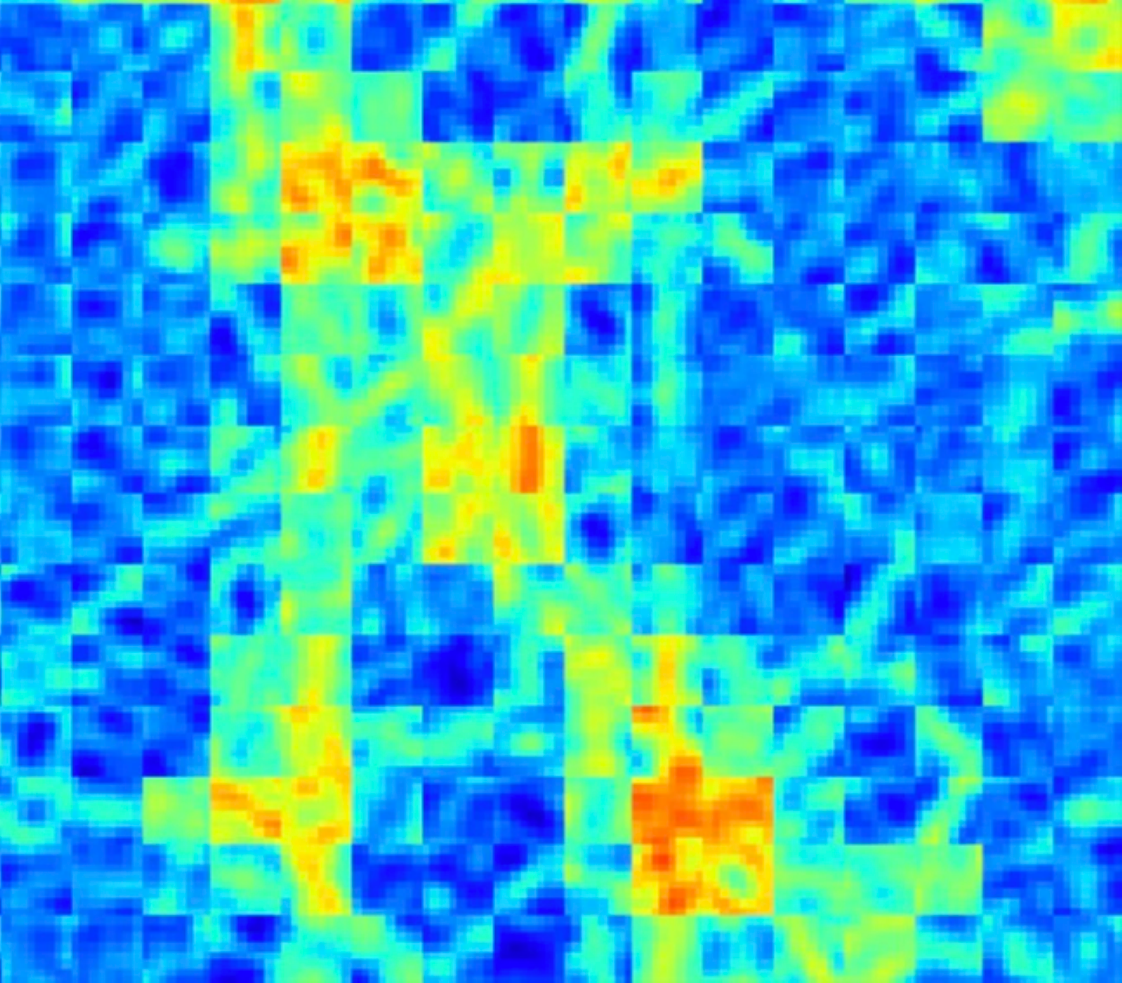} }}%
    \subfloat[\centering PtychoPINN]{{\includegraphics[height=4cm]{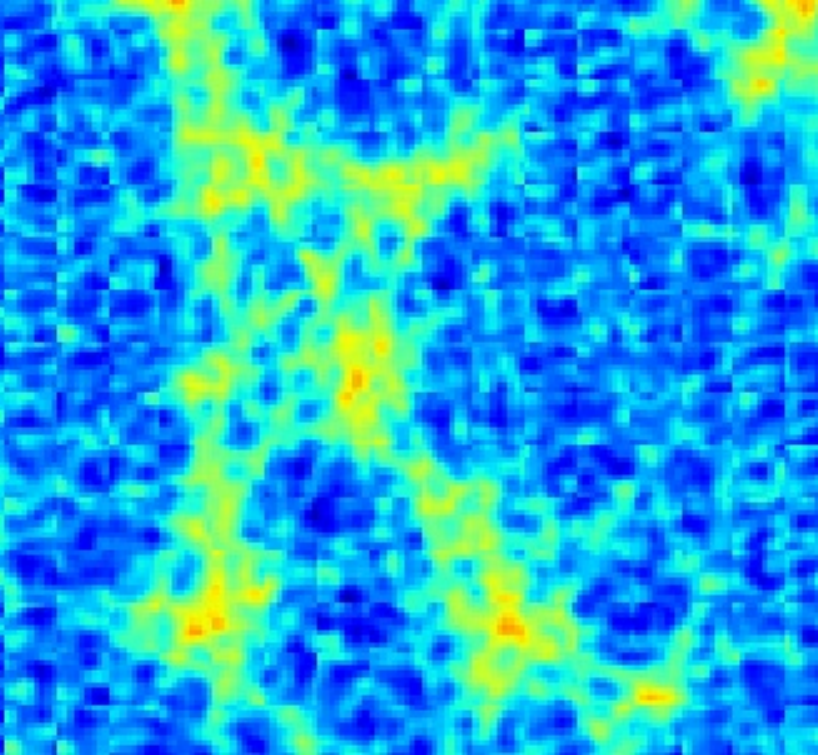} }}%
    
    \caption{\emph{Cross-training comparison.} Several models are first trained on locally-symmetric GRF objects, then reconstruct strongly-contrasting, asymmetric compositions of lines. The supervised model (b) struggles to reconstruct oriented features while the PINN -- both without (c) and with (d) ptychographic overlap constraints -- fares better.}
\label{fig:gen}%
\end{sidewaysfigure}

\begin{table}[h]
\begin{center}
\caption{Numerical study of out-of-distribution robustness corresponding to the data of Figure \ref{fig:gen_detailed}.}\label{tab2}%
\begin{tabular}{lcccc}
\toprule
           &      &      &  PSNR &  FRC50 \\
& Dataset & &       &        \\
\midrule
\multirow{4}{*}{PtychoPINN} & \multirow{2}{*}{train} & $A$ & 78.41 & 160 \\
           &      & $\phi$ & 69.11 & 158 \\
\cline{2-5}
           & \multirow{2}{*}{test} & $A$ & 61.11 &  42 \\
           &      & $\phi$ & 59.87 &  41 \\
\cline{1-5}
\cline{2-5}
\multirow{4}{*}{baseline} & \multirow{2}{*}{train} & $A$ & 73.72 &  76 \\
           &      & $\phi$ & 72.11 &  74 \\
\cline{2-5}
           & \multirow{2}{*}{test} & $A$ & 57.91 &  22 \\
           &      & $\phi$ & 55.77 &  18 \\
\bottomrule
\end{tabular}
\end{center}
\end{table}


\begin{sidewaysfigure}%
    \centering
    \subfloat[\centering Ground truth, training]{{\includegraphics[width=5cm]{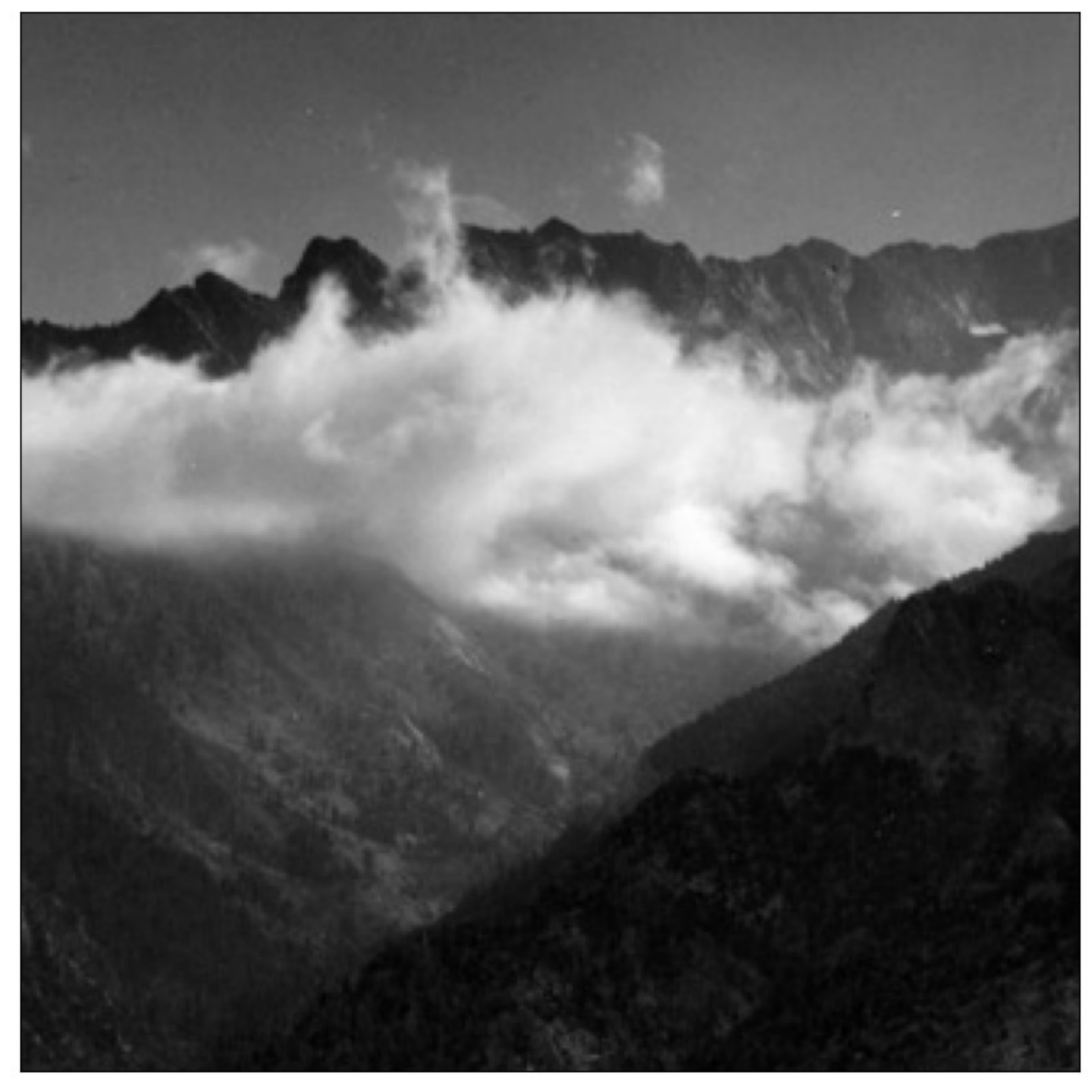} }}%
    \subfloat[\centering baseline (PtychoNN), in-sample ]{{\includegraphics[width=5cm]{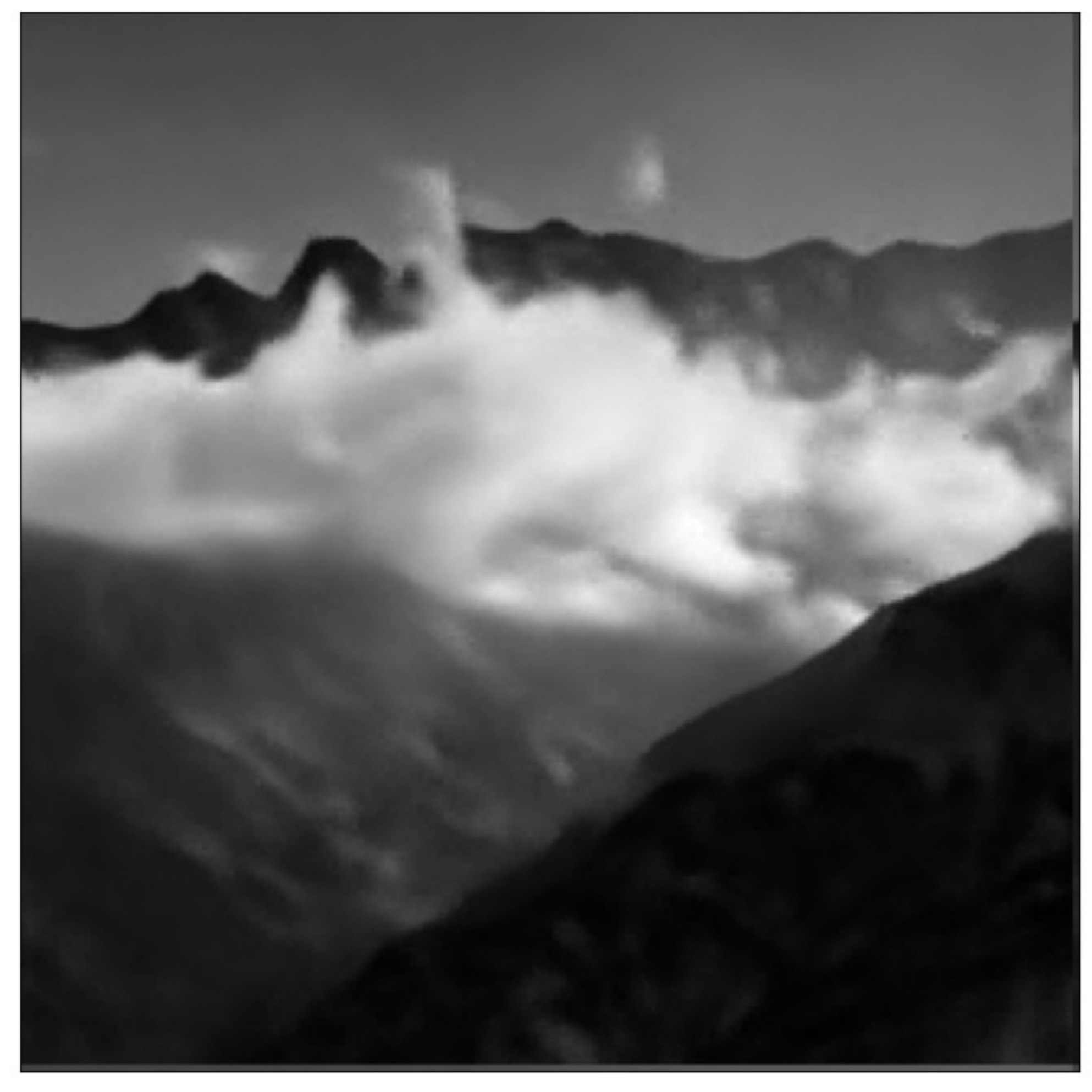} }}%
    \subfloat[\centering PtychoPINN, in-sample]{{\includegraphics[width=5cm]{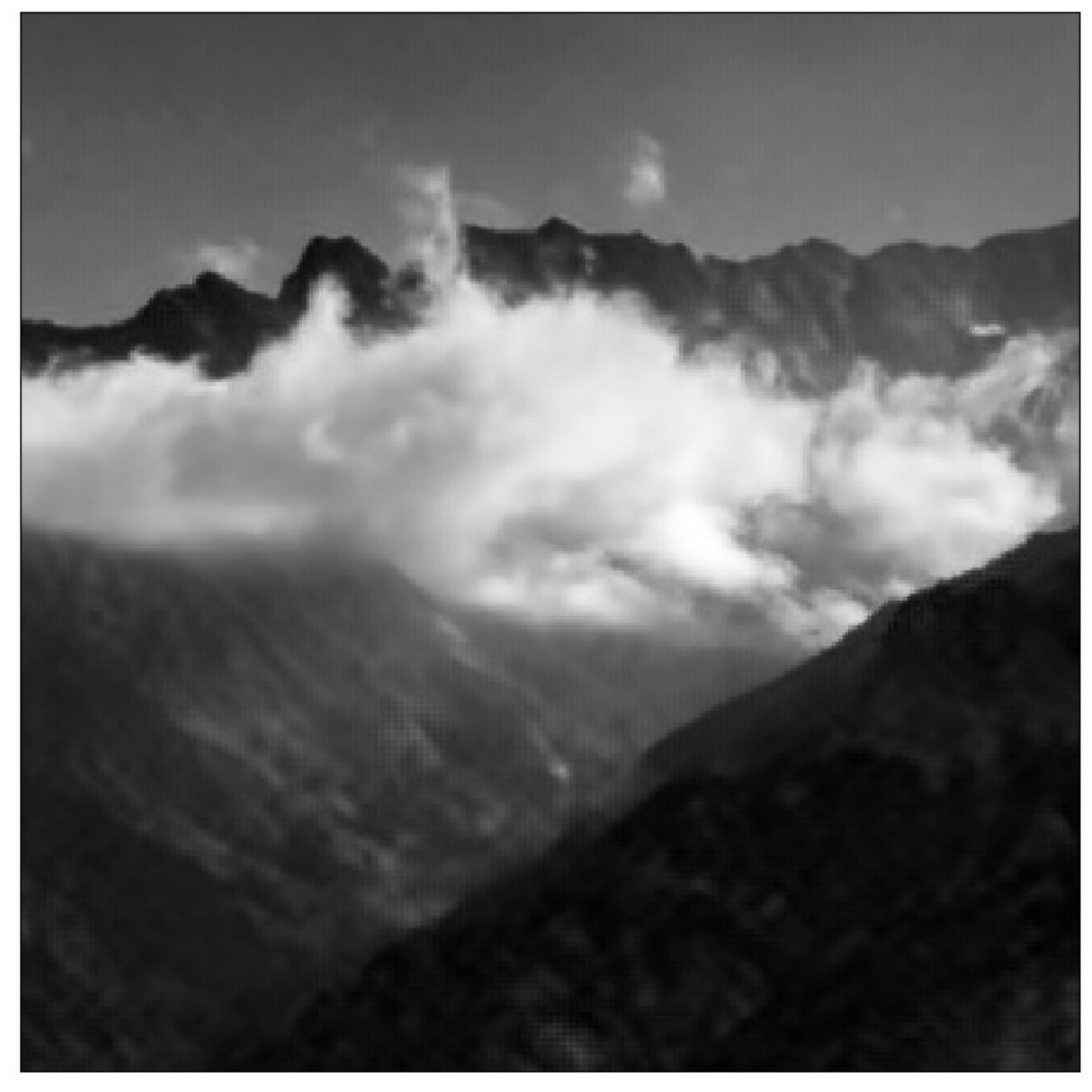} }}%
\vfill
    \subfloat[\centering Ground truth, out-of-distribution]{{\includegraphics[width=5cm]{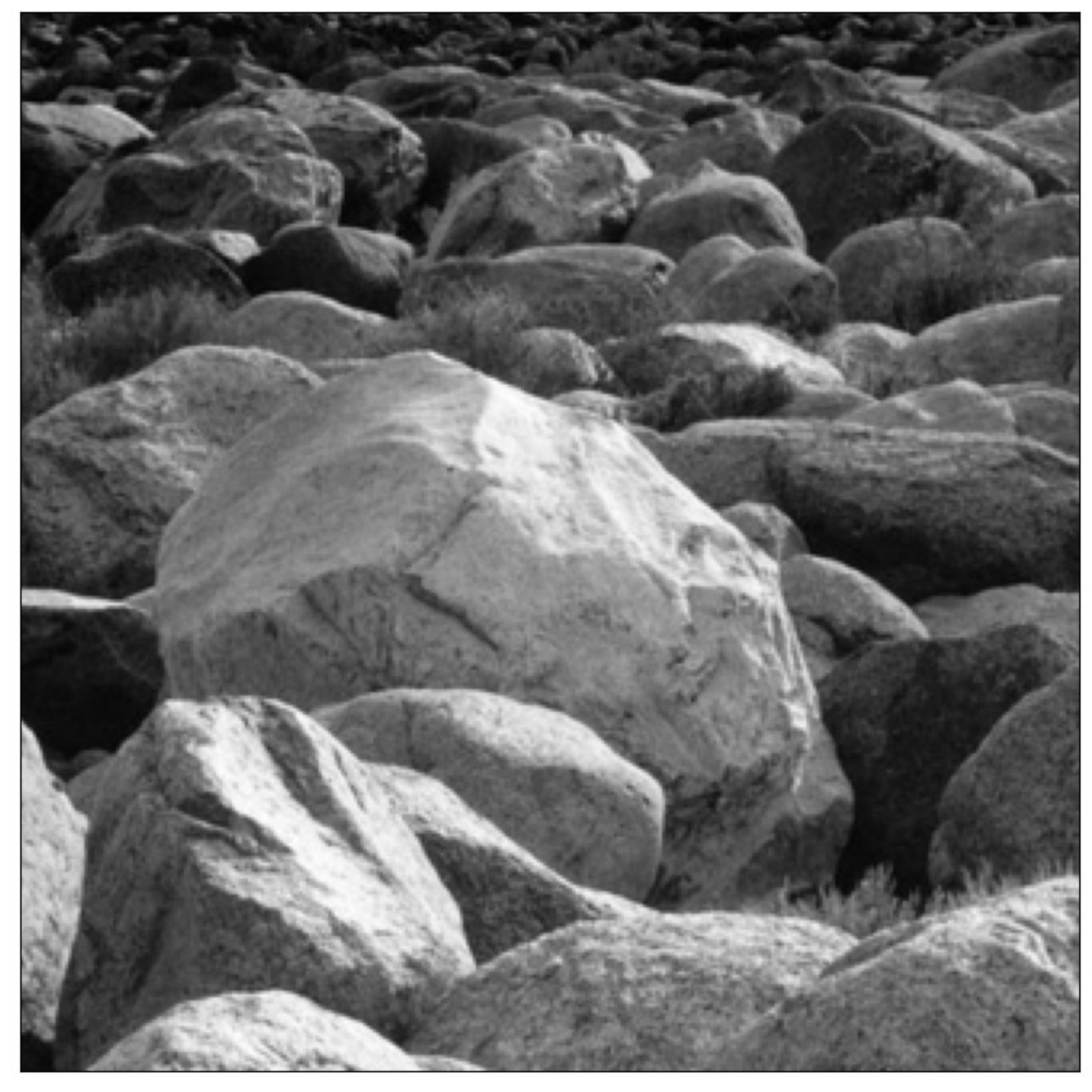} }}%
    \subfloat[\centering baseline (PtychoNN), out-of-distribution]{{\includegraphics[width=5cm]{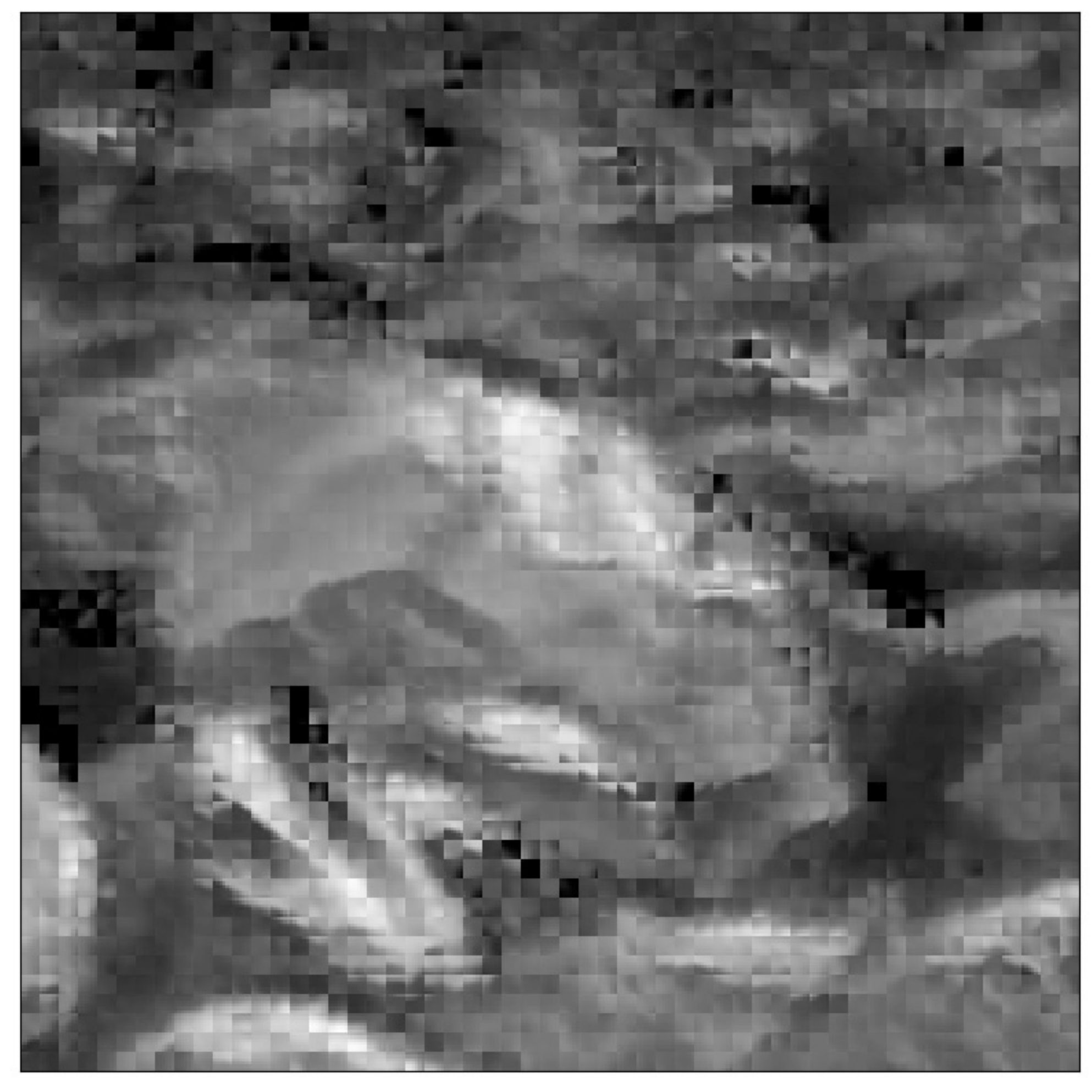} }}%
    \subfloat[\centering PtychoPINN, out-of-distribution]{{\includegraphics[width=5cm]{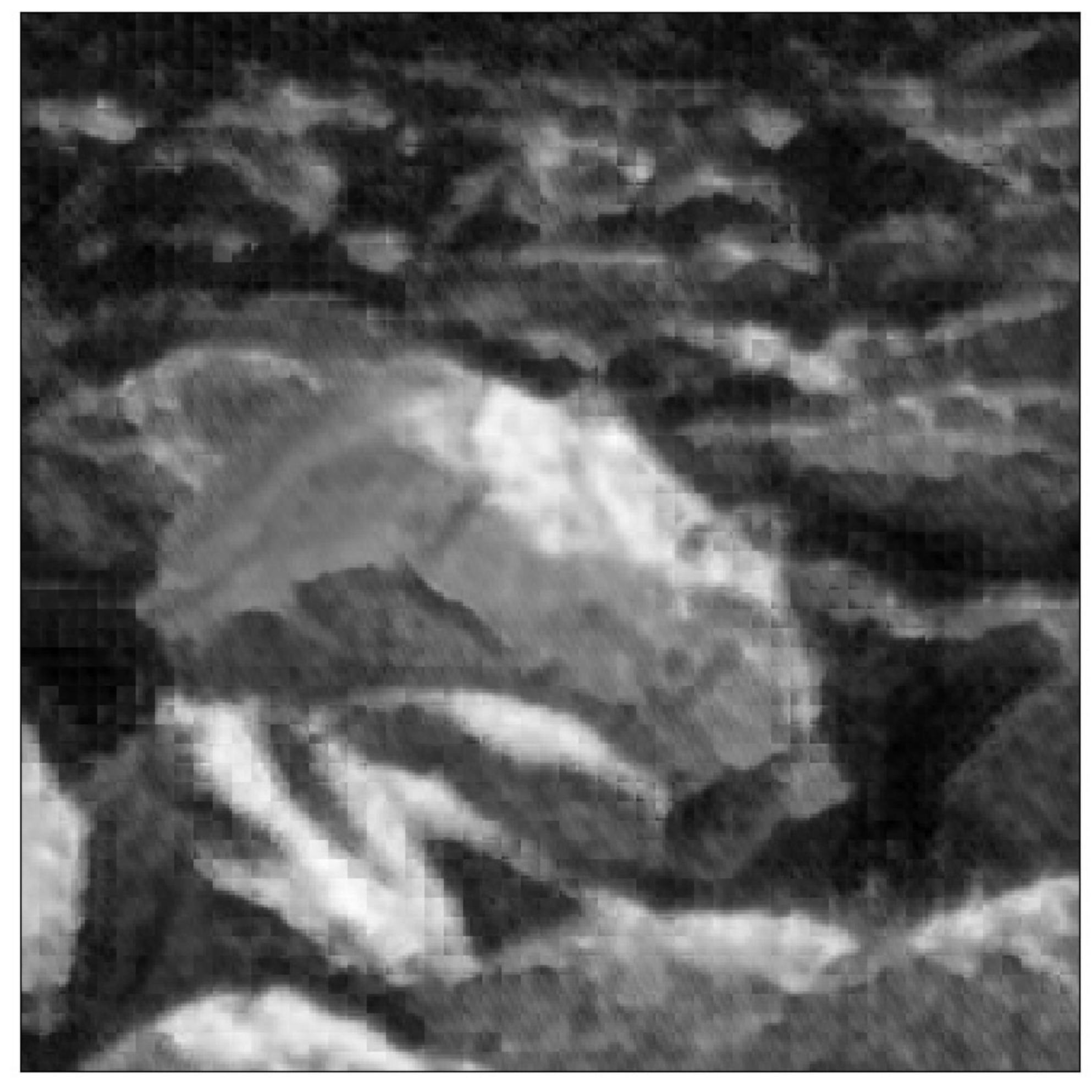} }}%
    \caption{\emph{Model Generalization.} This presents a comparison of the reconstructions produced by PtychoPINN and the baseline model, PtychoNN. Both models are trained using data derived from the amplitude object depicted in panel (a). Panels (b) and (c) show the training image as reconstructed by PtychoNN and PtychoPINN, respectively. The trained models then reconstruct diffraction from an out-of-distribution object (d), resulting in images (e) and (f) for PtychoNN and PtychoPINN, respectively. Both models degrade out-of-distribution, but PtychoPINN proves more robust. }%
    \label{fig:gen_detailed}%
\end{sidewaysfigure}

The full PINN model has another inherent advantage: superior generalizability. Even the basic PINN (that is, PtychoPINN stripped of overlap constraints) exhibits better out-of-distribution generalization compared to the supervised learning baseline in a qualitative out-of-distribution benchmark, shown in Figure \ref{fig:gen}. With full PtychoPINN, this advantage becomes more pronounced (Figure \ref{fig:gen} (d)).


From the point of view of the experimenter, the benefit of generalizability complements that of unsupervised training. An unsupervised architecture facilitates each instance of training; conversely, a more \emph{generalizable} model is more robust and does not need to be as frequently retrained. We attribute PtychoPINN's superior generalizability to the PINN structure, which necessitates the inverse map $G(X)$ to learn diffraction physics through its connection to the far-field diffraction map $F_d$ and diffraction-space loss function $L$. This connection enforces an approximation of physical consistency between the real-space reconstruction and the measured diffraction. Conversely, supervised training typically leads to violations of even basic conservation rules, such as the FT's unitarity (i.e., $\lVert x \rVert \approx \lVert F_c(G(x)) \rVert = \lVert F_d(\bar{y}) \rVert = \lVert \hat{x} \rVert$, by Parseval's identity and $\bar{y} \equiv F_c(G(x))$).

Furthermore, given that $\hat{x} = F_d(\bar{y})$ is not bijective and the loss function does not directly rely on $\bar{y}$, the PINN structure inherently overlooks variations in real-space structure that the diffraction measurement can't discern. Conversely, supervised training is susceptible to wasting model capacity by memorizing the training data's real-space structure. All told, supervised learning approaches face challenges in generalizability due to the difficult task of reconstructing an \emph{a priori} arbitrary map $X \rightarrow Y$ with less guidance by useful inductive biases.

As an aside, generalizability is closely connected to the concept of model \emph{rubustness}. Robustness depends on not only the network architecture but also the choice of loss function and properties of the data. Experimental data such as CDI images are often plagued by noise and distortions that must be modeled or explicitly removed for good results \cite{lee2021denoising}. While our framework takes a step in this direction by modeling experimental Poisson noise via its training objective, we leave a full treatment of the topic outside of the paper's scope.

\subsubsection{Resolution and Accuracy}

 The advances in reconstruction quality presented by PtychoPINN hold two-fold significance: they bear practical implications for scientific applications, and also offer potential insights into the workings of the model itself. To facilitate a more nuanced understanding of the model, we distinguish between the concepts of \emph{resolution} and \emph{accuracy}.

Resolution, while a critical component of good reconstruction, is not sufficient. By comparing $32 \times 32$ reconstruction patches from PtychoPINN, PtychoNN, and a basic PINN (Supplementary Figure 1), we observe that the basic PINN's reconstruction of line features displays a sharpness on par with PtychoPINN's. However, in contrast to both PtychoNN and PtychoPINN, the basic PINN misplaces these features. This arises due to the far field diffraction forward map's invariance to coordinate inversion.  In other words, real-space objects $O(r)$ and $O(-r)$ yield identical diffraction amplitudes. Thus, the basic PINN fails to distinguish between reconstruction pairs that are mirror images, or chiral, to each other.

It may be worthwhile here to make a distinction between local resolution - resolution within a single solution region - and overall resolution in an image that has been composed from multiple reconstruction patches. Local resolution pertains to the detail and sharpness within individual reconstruction patches, while reconstruction accuracy pertains to the entire stitched-together image. The latter requires correct $O(r) / O(-r)$ parity within individual reconstruction patches, a more stringent condition.The PINN architecture undoubtedly contributes a substantial enhancement to resolution, but it doesn't correspondingly boost \emph{accuracy}. The inclusion of real-space overlap constraints in PtychoPINN is needed to disambiguate between pairs of coordinate inversion-equivalent candidate objects, and leads to PtychoPINN's large enhancement in the accuracy of reconstruction.  However, this improvement primarily comes from the correct alignment of high-spatial frequency features, and not from an increase in the amount of information underlying individual images.

\section{Conclusion and Future Directions}
In conclusion, we present an autoencoder framework for scanning CDI into which we incorporate physical principles using PINN training and real-space constraints, thus gaining greatly improved accuracy, resolution, and generalizability compared to existing supervised deep learning-based lensless imaging methods. This new model, named PtychoPINN, inherits the intrinsic speed of NN approaches and is trained without labels. These attributes combine to make it a promising candidate for practical real-time, high-resolution imaging that transcends the resolution of lens-based systems without sacrificing imaging throughput. 

To train PtychoPINN, we incorporated a probabilistic (Poisson) model output and corresponding negative log likelihood (NLL) objective, thus modeling the Poisson noise intrinsic in experimental data. Our preliminary studies show that that the NLL objective significantly improves reconstruction quality in photon-limited scenarios. This suggests that selecting an appropriate loss function becomes increasingly critical in deep learning models that incorporate physics. In a similar vein, stochasticities such as probe jitter introduce uncertainties into the reconstruction that may degrade image quality. To mitigate this, we advocate for the exploration of probabilistic methods to improve performance and provide a principled estimation of model uncertainties. A more pressing omission in PtychoPINN is its lack of a mechanism for error inference in probe positioning—an important feature for robust experimental applications. To address this, we are currently developing this capability.

\backmatter

\section*{Declarations}

\begin{itemize}
\item This work was performed and partially supported by the US Department of Energy (DOE), Office of Science, Office of Basic Energy Sciences Data, Artificial Intelligence and Machine Learning at the DOE Scientific User Facilities program under the MLExchange Project (award No. 107514). Aashwin Mishra was partially supported by the SLAC ML Initiative.
\item The Authors declare no Competing Financial or Non-Financial Interests.
\item Acknowledgements: The authors would like to acknowledge insightful discussions with Daniel Ratner and Yousef Nashed. Additionally, we would like to acknowledge invaluable assistance and guidance from Matt Seaberg.
\item Code and Data Availability Statement: All data and code used in this study are available on GitHub. The repository, which includes both raw datasets and scripts for reproducing the simulations and analyses, can be accessed at github.com/hoidn/PtychoPINN. 
\item Author contributions: A.M. provided the project's initial suggestion and guided the overall scientific direction. O.H. proposed the approach, built the model, and performed the training and analyses. A.A.M. guided model development and elucidated connections to the scientific ML literature. O.H. and A.A.M. wrote the manuscript with input from A.M. All authors discussed results, edited the manuscript, and gave final approval for publication. 
\end{itemize}

\bigskip

\begin{appendices}






\end{appendices}


\bibliography{ptycho-pinn}


\end{document}


\renewcommand{\tablename}{Supplementary Table}
\renewcommand{\figurename}{Supplementary Figure}

\section{Supplemental materials}
\subsection{Tables and figures}
\begin{table}[h]
\begin{center}
\caption{Extension of Table 2 into a full ablation study encompassing the baseline supervised  model (PtychoNN, top row), PtychoPINN (bottom row), and two ablated versions of PtychoPINN, each containing one of the two defining features of the model (namely, ptychographic overlap constraints and the PINN/unsupervised structure).}\label{tab_ablation}
\begin{tabular}{p{2cm}l|ll|ll|ll}
\toprule
 & \multicolumn{1}{c}{} & \multicolumn{2}{c}{Lines} & \multicolumn{2}{c}{GRF} & \multicolumn{2}{c}{Large features}\\
\midrule
\bf{Feature set} & \bf{Metric}
& $\phi$ & $A$
& $\phi$ & $A$
& $\phi$ & $A$ \\
\midrule
$\{\}$\footnotemark[1]
& MAE & - & 0.201 & 0.0335 & 0.0153 & 0.219 & 0.0038 \\
& PSNR (dB) & - & 59.6 & 75.6 & 82.4 & 56.7 & 92.9 \\
& FRC50 ($\mathrm{pixel}^{-1}$) & - & 22.0 & 64.0 & 65.2 & 23.4 & 34.0 \\
\midrule
PINN
& MAE & - & 0.195 & 0.0859 & 0.0341 & 0.622 & 0.00581 \\
& PSNR (dB) & - & 59.7 & 67.5 & 75.4 & 50.3 & 88.8 \\
& FRC50 ($\mathrm{pixel}^{-1}$) & - & 22.0 & 29.7 & 64.7 & 8.2 & 13.9 \\
\midrule
overlaps
& MAE & - & 0.0755 & 0.0332 & 0.0158 & 0.187 & 0.00352 \\
& PSNR (dB) & - & 68.6 & 75.7 & 82.2 & 58.5 & 93.8 \\
& FRC50 ($\mathrm{pixel}^{-1}$) & - & 65.8 & 63.5 & 65.0 & 27.6 & 35.9 \\
\midrule
PINN,overlaps\footnotemark[2]
& MAE & - & \textbf{0.0473} & \textbf{0.0109} & \textbf{0.00507} & \textbf{0.149} & \textbf{0.00303} \\
& PSNR (dB) & - & \textbf{72.6} & \textbf{85.2} & \textbf{91.9} & \textbf{60.6} & \textbf{95.0} \\
& FRC50 ($\mathrm{pixel}^{-1}$) & - & \textbf{165.4} & \textbf{171.5} & \textbf{171.3} & \textbf{93.7} & \textbf{38.7} \\
\midrule
\end{tabular}
\end{center}
\footnotetext[1]{supervised baseline}
\footnotetext[2]{full PtychoPINN}
\end{table}

\begin{figure}
    \centering
    \subfloat{{\includegraphics[width=6.5cm]{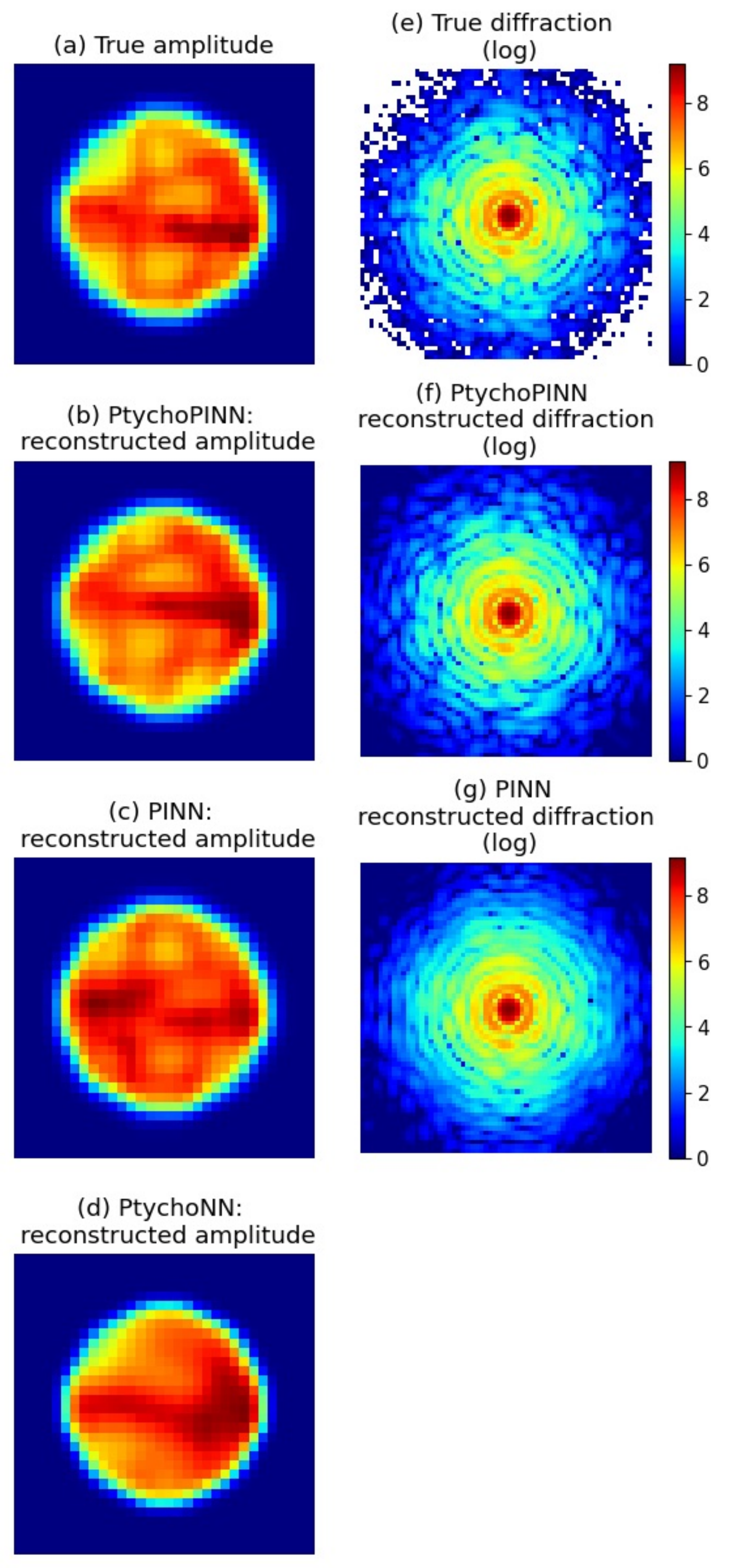} }}
    \caption{\emph{PINN and parity}. Comparison of real-space and diffraction reconstructions from PtychoPINN (b) and the basic PINN (c) with no overlap constraints. Note that both PtychoPINN and the basic PINN reconstruct small features, but only PtychoPINN resolved the inversion degeneracy correctly. The supervised-training baseline (d) produces a reconstruction that has the correct asymmetry, but is considerably blurred.}%
    \label{fig:patches}
\end{figure}

\subsection{Plotting details}
All amplitudes images are plotted with an auto-scaled color map. Because the model introduces a training run-specific normalization factor into reconstructed amplitudes, we omit numerical scales in the presentation of amplitude images.